\documentclass[amsmath, amssymb, superscriptaddress, reprint]{revtex4-1}
\usepackage{graphicx}
\usepackage{dcolumn}
\usepackage{bm}
\usepackage[unicode=true,pdfusetitle,
 bookmarks=true,bookmarksnumbered=false,bookmarksopen=false,
 breaklinks=false,pdfborder={0 0 1},backref=false,colorlinks=false]
 {hyperref}
\usepackage{placeins}
\usepackage{float}

\begin{document}

\preprint{AIP/123-QED}

\title[Magnon-Phonon Thermometry]{Magnon-Phonon Quantum Correlation Thermometry}

\author{C.A. Potts}
\email{cpotts@ualberta.ca}
\affiliation{Department of Physics, University of Alberta, Edmonton, Alberta T6G 2E9, Canada}

\author{V.A.S.V. Bittencourt}%
\affiliation{Max Planck Institute for the Science of Light, Staudtstr. 2, PLZ 91058 Erlangen, Germany}%

\author{S. {Viola Kusminskiy}}
\affiliation{Max Planck Institute for the Science of Light, Staudtstr. 2, PLZ 91058 Erlangen, Germany}%
\affiliation{Institute for Theoretical Physics, University Erlangen-Nuremberg, Staudtstr. 7, PLZ 91058 Erlangen, Germany}

\author{J.P. Davis}
\email{jdavis@ualberta.ca}
\affiliation{Department of Physics, University of Alberta, Edmonton, Alberta T6G 2E9, Canada}

\begin{abstract}

A large fraction of quantum science and technology requires low-temperature environments such as those afforded by dilution refrigerators. In these cryogenic environments, accurate thermometry can be difficult to implement, expensive, and often requires calibration to an external reference. Here, we theoretically propose a primary thermometer based on measurement of a hybrid system consisting of phonons coupled via a magnetostrictive interaction to magnons. Thermometry is based on a cross-correlation measurement in which the spectrum of back-action driven motion is used to scale the thermomechanical motion, providing a direct measurement of the phonon temperature independent of experimental parameters.  Combined with a simple low-temperature compatible microwave cavity read-out, this primary thermometer is expected to become a popular thermometer for experiments below 1 K.  

\end{abstract}

\pacs{Valid PACS appear here}
\maketitle

\section{Introduction}
Achieving low temperatures is a central aspect of many fields of physics, from the study of phase transitions in condensed matter physics, to accessing quantum behavior for fundamental tests and emerging technologies.  Hand in hand with the development of techniques for achieving low temperatures is the measurement of these temperatures.  While many physical systems demonstrate temperature dependent behavior that can be used for thermometry, such as electrical resistance or magnetic susceptibility, such thermometers rely on extrinsic properties and therefore require calibration to an external reference to be of use \cite{yeager_thermometry_2001}.  These types of thermometers are referred to as secondary thermometers.  Another class of thermometers, called primary thermometers, instead do not require external calibration and are therefore critical to precision measurements and temperature metrology \cite{Preston-Thomas_International_1989, Rusby_Provisional_2002, Rusby_Realization_2007}.  Primary thermometers used in cryogenic experiments include those that depend on intrinsic physical properties such as the vapor pressure of $^3$He or $^4$He \cite{Preston-Thomas_International_1989}, the melting curve of $^3$He \cite{greywall_3He_1982}, and orientation of radiation from the decay of radioactive $^{60}$Co \cite{berglund_nuclear_1972}.  Unfortunately these primary thermometers are often impractical for a variety of reasons, such as high cost or incompatibility with experimental apparatus.  For example, the melting pressure thermometry of $^3$He---which forms the basis for the realization of the international temperature scale below 1 K \cite{greywall_3He_1982, Rusby_Realization_2007}---requires difficult design and construction \cite{Pollanen_Resistance_2009}, in addition to rare and expensive $^3$He, and the scintillators used to measure nuclear orientation thermometers are incompatible with applied magnetic fields.  

An alternative group of primary thermometers is based on the principle of measuring intrinsic thermal noise.  While most efforts in noise thermometry have focused on electrical noise in resistors \cite{Johnson_Thermal_1928, Nyquist_Thermal_1928, Kamper_Noise_1971, rothfuss_Noise_2013, Rothfuss_Noise_2016, Shibahara_Primary_2016}, a promising new candidate comes from thermomechanical noise \cite{Hauer_General_2013} since cavity-optomechanics has enabled thermomechanical noise to be the dominant noise source (instead of measurement noise) even at millikelvin temperatures \cite{aspelmeyer_cavity_2014}.  While it has been shown that thermomechanical noise can be used for noise thermometry when the measurement is calibrated using an external modulation tone \cite{MacDonald_Optomechanics_2016, Gorodetksy_Phase_2010}, this constitutes a secondary thermometer.  Instead, primary thermometry can be achieved via self-calibration by comparing the thermomechanical signal with intrinsic quantum fluctuations \cite{borkje_observability_2010,purdy_quantum_2017}.  To date, this quantum correlation thermometry has been measured with cavity optomechanics down to 10 K  \cite{purdy_quantum_2017}, but the need for primary thermometers lies below 1 K.  Unfortunately, heating of optomechanical resonators from optical absorption has been found to be a major problem at such low temperatures \cite{Meenehan_Pulsed_2015,hauer_two-level_2018,ramp_elimination_2019}, calling into question the practicality of optomechanical noise thermometry.  

\begin{figure}[t]
\includegraphics[width=7.5 cm]{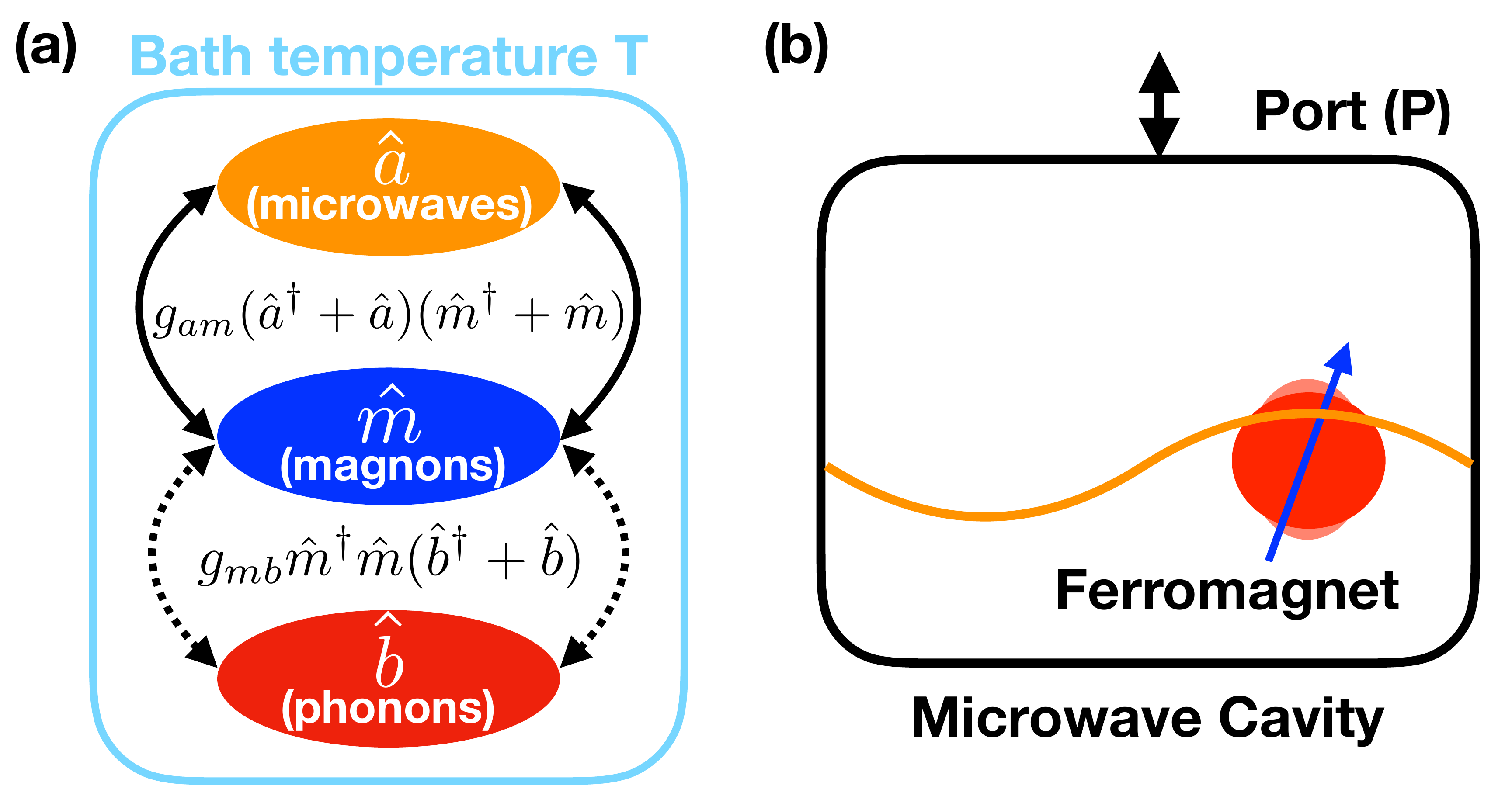}
\caption{Considered setup. (a) A single mode of a microwave cavity ($\hat{a}$) couples linearly to the magnetic excitations  ($\hat{m}$) of a magnetic material loaded in the cavity. The later are coupled parametrically to the mechanical vibrations of the material ($\hat{b}$). 
(b) Schematic of a possible experimental implementation: the microwave mode couples to the uniform magnon mode which in turn is coupled to the vibrations of the material. The cavity can be externally pumped via a port $P$. Our scheme consists of pumping the microwave mode and measuring noise correlations via the port $P$, which carry information about the bath temperature $T$.}
\label{Fig:01}
\end{figure}

Luckily, cavity electromechanics provides similar measurement capabilities to cavity optomechanics, but with orders of magnitude lower energy photons and dissipationless superconducting materials \cite{Woolley_Nanomechanical_2008,Regal_Measuring_2008}. Therefore, we are motivated to explore quantum correlation primary noise thermometry of mechanical resonators using cavity electromechanics.  While nanofabricated on-chip electromechanical systems have proven to be a powerful platform for fundamental demonstrations of quantum mechanics \cite{Palomaki_Entangling_2013,Suh_Mechanically_2014,Pirkkalainen2015} and quantum technology \cite{Lecocq_Mechanically_2016}, a simpler platform is desirable for wide adoption as a thermometer.  Hybrid magnon-phonon systems provide just such a platform, with the relative ease with which magnons can be strongly coupled to microwave cavities \cite{zhang_strongly_2014,goryachev_high_coop_2014,tabuchi_hybridizing_2014}, the commercial availability of high-quality materials, and the fact that magnons have been shown to couple to phonons in dielectric magnetic spheres \cite{zhang_cavity_2016}. Therefore, in this article, we propose and theoretically explore a primary thermometer for a coupled microwave-magnon-photon system, depicted in Fig.~\ref{Fig:01}.  This particular setup is inspired by recent developments in microwave cavity systems with an embedded low-loss ferromagnetic element \cite{zhang_strongly_2014, tabuchi_hybridizing_2014, goryachev_high_coop_2014, tabuchi_coherent_coupling_2015, bai_spin_pumping_2015, viennot_coherent_coupling_2015, zhang_cavity_quatum_2015, cao_exchange_2015, zhang_magnon_dark_2015, hisatomi_bidirectional_conversion_2016, lachance_resolving_quanta_2017, wang_bistability_2018, hou_strong_coupling_2019,Lachance_Hybrid_2019,Wang_Nonreciprocity_2019,zhang_cavity_2016}. 

In this arrangement, a single mode of a microwave (MW) cavity couples to the magnetic excitations (magnons) of a magnetic material via a linear interaction. The magnons in turn interact with a vibrational mode of the material via the magnetostrictive interaction \cite{zhang_cavity_2016}. To describe the cryogenic environment in which these experiments are carried out, we assume that each mode interacts with a corresponding heat bath and that all the baths are at the same temperature $T$.  The aim is then to measure the temperature of the phonon heat bath by pumping and measuring the output noise fluctuations of the MW cavity.

\section{Model}

We consider the hybrid cavity microwave-magnon-phonon system depicted in Fig.~\ref{Fig:01}, which we describe in terms of three coupled bosonic modes, denoted by $\hat{a}$ (cavity microwave mode), $\hat{m}$ (magnon mode) and $\hat{b}$ (phonon mode), with frequencies $\omega_{\rm{a}}$, $\omega_{\rm{m}}$ and $\omega_{\rm{b}}$ respectively. We assume that the MW and magnon modes interact via a linear coupling Hamiltonian while the magnons and the mechanical vibrations are coupled via a parametric type Hamiltonian (see Fig.~\ref{Fig:01}a), such that the total Hamiltonian describing the system dynamics reads
\begin{equation}
\begin{aligned}
    \hat{\mathcal{H}}_{0} &= \hbar\omega_{\rm{a}}\hat{a}^{\dagger}\hat{a} + \hbar\omega_{\rm{b}}\hat{b}^{\dagger}\hat{b}+\hbar\omega_{\rm{m}}\hat{m}^{\dagger}\hat{m} \\&+\hbar g_{\rm{am}}(\hat{a}+\hat{a}^{\dagger})(\hat{m}+\hat{m}^{\dagger}) + \hbar g_{\rm{mb}}\hat{m}^{\dagger}\hat{m}(\hat{b}+\hat{b}^{\dagger}).
\label{eqn:Ham}
\end{aligned}
\end{equation}
Here, the MW-magnon coupling strength is indicated by $g_{\rm{am}}$ while the magnon-phonon coupling strength is $g_{\rm{mb}}$. The Hamiltonian, Eq.~\eqref{eqn:Ham}, describes a general microwave-magnon-phonon system. This model applies directly to current experimental setups in which the MW mode strongly couples with the uniform magnetization mode (Kittel mode) of a ferromagnetic yttrium-iron garnet (YIG) sphere \cite{zhang_strongly_2014,goryachev_high_coop_2014, tabuchi_hybridizing_2014, zhang_cavity_2016}. The resonant MW-magnon coupling is usually realized by tuning the frequency of the Kittel mode by an applied external DC magnetic field (35 to 350 mT), with frequencies ranging between 1 to 10 GHz. Magnetoelastic effects are responsible for the coupling between magnons and phonons corresponding to the collective mechanical breathing modes of the YIG sphere, usually in the MHz range due to the relative large size of the sphere (currently in the 100 $\mu$m radius range). For this standard experimental setup, the coupling strengths are $g_{\rm{am}} \sim 10 $ MHz and $g_{\rm{mb}} \sim 10$ mHz \cite{zhang_cavity_2016}.  In general, this model can describe more complex structures and/or modes, with coupling parameters modified accordingly.

The MW cavity is assumed to be coupled to a single external port, labeled $P$, which is coherently driven such that the total Hamiltonian of the system is
\begin{equation}
\hat{\mathcal{H}} = \hat{\mathcal{H}}_{0} + \hat{\mathcal{H}}_{\rm{drive}},
\label{eq:Ham02}
\end{equation}
where $\hat{\mathcal{H}}_{\rm{drive}}=i\hbar\epsilon_{\rm{d}}\sqrt{\kappa_{\rm{P}} }(\hat{a}e^{i\omega_{\rm{d}} t} - \hat{a}^{\dagger}e^{-i\omega_{\rm{d}} t})$, with $\kappa_{\rm{P}}$ the coupling rate to the port and $\omega_{\rm{d}}$ the driving frequency. The protocol that follows can be also implemented in multiple port setups via cross-correlation measures, although the single port approach has some experimental advantages. Namely, to avoid errors, the coupling rate to both ports must be matched exactly, which can be difficult to implement with two or more ports.

Employing standard quantum optics procedures, we obtain the linearized Hamiltonian for the fluctuations $\delta \hat{A} = \hat{A} - \langle \hat{A} \rangle$ around the steady state value $\langle \hat{A} \rangle$ of the fields $\hat{A} = \hat{a},\hat{m}, \hat{b}$ (see Appendix A). We further assume that $g_{\rm{am}} \gg g_{\rm{mb}}$ and that the photon and magnon modes are resonant. In the frame rotating at the drive frequency, and considering the rotating wave approximation for the magnon-photon coupling, the linearized Hamiltonian reads
\begin{equation}
\begin{aligned}
    \hat{\mathcal{H}}_{\rm{Lin}} &= -\hbar\Delta_{\rm{a}}\delta\hat{a}^{\dagger}\delta\hat{a} + \hbar\omega_{\rm{b}}\delta\hat{b}^{\dagger}\delta\hat{b} -\hbar \tilde{\Delta}_{\rm{m}}\delta\hat{m}^{\dagger}\delta\hat{m}\\&
    +\hbar g_{\rm{am}}(\delta\hat{a}\delta\hat{m}^{\dagger} + \delta\hat{a}^{\dagger}\delta\hat{m}) 
    \\&+\hbar (G_{\rm{mb}}\delta\hat{m}^{\dagger}+G_{\rm{mb}}^*\delta\hat{m}) ( \delta \hat{b} + \delta \hat{b}^\dagger ),
\label{eqn:LinHam}
\end{aligned}
\end{equation}

\noindent where $\Delta_{\rm{a}} = \omega_{\rm{d}}-\omega_{\rm{a}}$ and $\tilde{\Delta}_{\rm{m}}=\omega_{\rm{d}}-\omega_{\rm{m}} - 2 \hbar g_{\rm{mb}} \mbox{Re}[\langle \hat{b} \rangle]$ denote the detuning of the drive with respect to the bare microwave cavity and the phonon-shifted magnon frequencies respectively (see Appendix A). Since the magnon-phonon coupling is the lowest rate in the system, in the following we do not consider the frequency shift $- 2 \hbar g_{\rm{mb}} \mbox{Re}[\langle \hat{b} \rangle]$. The effective magnon-phonon coupling is defined as $G_{\rm{mb}} = g_{\rm{mb}} \langle \hat{m} \rangle$ and is therefore enhanced from its bare value $g_{\rm{mb}}$ by the average number of steady state magnons, driven via the coupling to the MW mode.

From the Hamiltonian, we derive the quantum Langevin equations of motion for the frequency domain operators $\delta\hat{\mathcal{O}}[\omega] = \int_{-\infty}^{\infty} dt e^{-i\omega t} \delta\hat{\mathcal{O}}(t)$: 
\begin{equation}
\begin{aligned}
    \chi_{\rm{a}}^{-1}[\omega]\delta\hat{a}[\omega] &= - i g_{\rm{am}}\delta\hat{m}[\omega] + \sqrt{\kappa_{\rm{P}}} \hat{\xi}_{\rm{P}}[\omega], \\
    \chi_{\rm{m}}^{-1}[\omega]\delta\hat{m}[\omega] &= -i g_{\rm{am}}\delta\hat{a}[\omega] - iG_{\rm{mb}}\delta\hat{z}[\omega]  + \sqrt{\gamma_{\rm{m}}} \hat{\eta}[\omega],\\
    \delta\hat{z}[\omega] &= -i(\chi_{\rm{b}}[\omega]-\chi_{\rm{b}}^*[-\omega])\times\\
    &\Big[ G_{\rm{mb}}\delta\hat{m}^{\dagger}[-\omega]+G_{\rm{mb}}^*\delta\hat{m}[\omega] + \delta \hat{F}_{\rm{th}}[\omega] \Big],
\label{eqn:EOM}
\end{aligned}
\end{equation}

\noindent where $\chi_{\rm{a}}(\omega) = [-i(\Delta_{\rm{a}} + \omega) + \kappa/2]^{-1}$, $\chi_{\rm{m}}(\omega) = [-i(\tilde{\Delta}_{\rm{m}} + \omega) + \gamma_{\rm{m}}/2]^{-1}$ and $\chi_{\rm{b}}(\omega) = [i(\omega_{\rm{b}} - \omega) + \gamma_{\rm{b}}/2]^{-1}$ are respectively the photon, magnon and phonon susceptibilities, with $\gamma_{\rm{m(b)}}$ the magnon (phonon) decay rate.  The total cavity mode decay rate $\kappa = \kappa_{\rm{P}} + \kappa_{\rm{I}}$ includes the decay into the $P$ channel as well as the intrinsic decay rate $\kappa_{\rm{I}}$.  In the last equation we have defined the phonon displacement operator $\delta \hat{z}[\omega] = \delta \hat{b}[\omega] + \delta \hat{b}^\dagger [\omega]$. 

The open dynamics of the system are described via input fluctuation operators. The input fluctuations of the cavity mode are denoted by $\hat{\xi}_{\rm{P}}[\omega]$, and for the magnon mode by $\hat{\eta}[\omega]$, whereas the noise acting on the phonon mode is denoted by $\delta \hat{F}_{\rm{th}}[\omega]$. These operators have correlations satisfying the fluctuation-dissipation theorem \cite{callen_irreversibility_1951, kubo_the_fluctuation_1966, gardiner_quantum_2000}. We assume that the magnon, photon and phonon environments are heat baths that all have the same temperature $T$. For describing the photon and magnon environments, we use the standard framework of the first Markov approximation (the environment correlations decay much faster than the time scale in which the system has a considerable evolution) and consider that the initial system-bath state is uncorrelated \cite{gardiner_quantum_2000, breuer_the_theory_2002}. We moreover assume that the state of each environment is weakly affected by the system and is described by thermal states. The correlation properties of the magnon and MW noises $\hat{\beta}=\hat{\eta},\hat{\xi}_{\rm{P}}$, are then given by
\begin{equation}
\begin{aligned}
\langle \hat{\beta}[\omega] \hat{\beta}^\dagger[\omega^\prime] \rangle &=  2 \pi(n_{\rm{th}}+1)\delta(\omega + \omega^{\prime}), \\
\langle \hat{\beta}^\dagger[\omega] \hat{\beta}[\omega^\prime] \rangle &= 2 \pi n_{\rm{th}}\delta(\omega + \omega^{\prime}),
\label{eqn:PhotonNoise}
\end{aligned}
\end{equation}
where $n_{\rm{th}} = \left[{\rm{exp}}(\hbar \omega_{\rm{a,m}}/k_{\rm{B}} T) -1 \right]^{-1}$ is the thermal occupancy of the photonic and magnonic baths. Since the magnon and photon modes are assumed to be resonant ($\omega_{\rm{a}} = \omega_{\rm{m}}$), $n_{\rm{th}}$ is identical for both $\hat{\eta}$ and $\hat{\xi}_{\rm{P}}$.

The noise acting on the phonon mode is encoded in $\delta\hat{F}_{\rm{th}}[\omega]$, which represents the effects of the environment on the phonon mode and is given by \cite{giovannetti_phase_2001} (see also the appendix B) 
\begin{equation}
    \int_{-\infty}^{\infty} d \omega^\prime \langle \{ \delta \hat{F}_{\rm{th}}[\omega^{\prime}],\delta \hat{F}_{\rm{th}}[\omega] \} \rangle =  2 \pi  \gamma_{\rm{b}} \frac{\omega}{\omega_{\rm{b}}} {\rm{coth}} \bigg( \frac{\hbar\omega}{2k_{\rm{B}} T} \bigg),
\label{eqn:ThermalSpectrum}
\end{equation}
where $\{\cdot, \cdot \}$ represents the anti-commutator. This model is in correspondence with the thermomechanical model for phonon modes in cavity optomechanical systems \cite{giovannetti_phase_2001, purdy_quantum_2017}. The symmetrized noise spectra is required to compare with the experimentally observable correlation functions \cite{giovannetti_phase_2001,clerk_introduction_2010}. Note that although we have used a colored-noise model for the phonon mode, the magnon mode noise is white. This is a good approximation at low temperatures, such that the number of thermal magnon excitations is small, and the magnon mode quality factor is large (see the discussion in Ref.~\cite{giovannetti_phase_2001}).

The Langevin equations, Eq.~\eqref{eqn:EOM}, are then  solved. The cavity field fluctuations are given in terms of $\delta \hat{z}$ by (see Appendix B)
\begin{equation}
\begin{aligned}
    \delta\hat{a}[\omega] &= -\Lambda_{\rm{am}}[\omega](g_{\rm{am}}G_{\rm{mb}}\chi_{\rm{m}}[\omega]\delta\hat{z}[\omega]+
    \\&i g_{\rm{am}}\chi_{\rm{m}}[\omega]\sqrt{\gamma_{\rm{m}}}\delta\hat{\eta}[\omega]-\sqrt{\kappa_{\rm{P}}}\hat{\xi}_{\rm{P}}[\omega]),
\label{eqn:Deltaa}
\end{aligned}
\end{equation}
\noindent with $\Lambda_{\rm{am}}[\omega] = \big[\chi_{\rm{a}}^{-1}[\omega]+g_{\rm{am}}^2\chi_{\rm{m}}[\omega]\big]^{-1}$. The thermal-mechanical fluctuations, encoded in $\delta \hat{z}[\omega]$, are imprinted on the microwave mode via the coupling to the magnon mode. This is akin to the cavity optomechanical case, in which the thermal phonon fluctuations can be measured via the noise of an optical mode \cite{aspelmeyer_cavity_2014,purdy_quantum_2017}.

\section{Noise spectrum and thermometry of the mechanical vibrations}

Experimentally, the microwave modes are only accessible via the reflected or transmitted signals. We can obtain the fluctuations of the output mode via the input-output relation \cite{gardiner_quantum_2000}
\begin{equation}
    \delta\hat{a}_{{\rm{out}}}[\omega] = \hat{\xi}_{\rm{P}}[\omega] - \sqrt{\kappa_{\rm{P}}}\delta\hat{a}[\omega].
\label{eqn:aout}
\end{equation}
Detection schemes, such as homodyne, can measure arbitrary quadratures of the output fields. These carry information of the phase and amplitude fluctuations and are affected by thermal noise \cite{clerk_introduction_2010}. Here we use the canonical in-phase and out-of-phase quadratures to construct correlation spectra.  For an arbitrary operator $\hat{O}$ we define $\hat{X}_{\hat{O}}[\omega] = \hat{O}[\omega] +\hat{O}^\dagger[-\omega]$, and $\hat{Y}_{\hat{O}}[\omega] = -i\left(\hat{O}[\omega] - \hat{O}^\dagger[-\omega]\right)$. At zero detuning $\Delta_{\rm{a}} =\Delta_{\rm{m}}= 0$ the quadratures of the cavity field are given by

\begin{widetext}
\begin{equation}
\begin{aligned}
    \hat{X}_{\delta \hat{a}}[\omega] &= \sqrt{\gamma_{\rm{m}}}\Lambda_{\rm{am}}[\omega]g_{\rm{am}}\chi_{\rm{m}}[\omega] \hat{Y}_{\hat{\eta}}[\omega]+\Lambda_{\rm{am}}[\omega] \hat{X}_{\hat{\xi}}[\omega],\\
    \hat{Y}_{\delta \hat{a}}[\omega] &=  2\Lambda_{\rm{am}}[\omega] \vert G_{\rm{mb}} \vert^2 g_{\rm{am}} \chi_{\rm{m}}[\omega] \big(\sqrt{\gamma_{\rm{m}}} f_{\rm{m}} [\omega] \hat{Y}_{\hat{\eta}}[\omega]-f_{\rm{a}}[\omega] \delta\hat{X}_{\hat{\xi}}[\omega] \big) -\sqrt{\gamma_{\rm{m}}} \Lambda_{\rm{am}}[\omega] g_{\rm{am}} \chi_{\rm{m}}[\omega] \hat{X}_{\hat{\eta}}[\omega] \\ &\quad +\Lambda_{\rm{am}}[\omega] \hat{Y}_{\hat{\xi}}[\omega] +2i(\chi_{\rm{b}}[\omega]-\chi_{\rm{b}}^*[-\omega])\Lambda_{\rm{am}}[\omega]\vert G_{\rm{mb}} \vert g_{\rm{am}}\chi_{\rm{m}}[\omega]\delta\hat{F}_{\rm{th}}[\omega],
\label{eqn:Quada}
\end{aligned}
\end{equation}
\end{widetext}
where the coefficients $f_{\rm{a}}[\omega]$ and $f_{\rm{m}}[\omega]$ have been defined in Appendix B and we have adopted the short hand notation for the total MW input noise $\hat{\xi} = \sqrt{\kappa_{\rm{P}}} \hat{\xi}_{\rm{P}}$. 

Using these and Eq.~\eqref{eqn:aout} we can construct a generic quadrature of the output field $\delta \hat{a}_{{\rm{out}} }$ parameterized by $\theta$, as
\begin{equation}
    \hat{X}_{\delta\hat{a}_{{\rm{out}}},\theta}[\omega]= \cos{(\theta)}  \hat{X}_{\delta\hat{a}_{{\rm{out}}}}[\omega] + \sin{(\theta)} \hat{Y}_{\delta\hat{a}_{{\rm{out}}}}[\omega],
\label{eqn:ArbQuad}
\end{equation}
such that the symmetrized correlation spectrum can be calculated as
\begin{equation}
    S_{\theta, \theta^\prime}[\omega]=\frac{1}{4} \displaystyle \int_{-\infty}^\infty d\omega^\prime \langle \{  \hat{X}_{\delta\hat{a}_{{\rm{out}}},\theta}[\omega], \hat{X}_{\delta\hat{a}_{{\rm{out}}},\theta^\prime}[\omega^\prime] \} \rangle.
\label{eqn:NoiseSpec}
\end{equation}
The reflected signal can be demodulated using an IQ-mixer allowing the simultaneous measurement of $\delta \hat{X}_{\delta\hat{a}_{{\rm{out}}}}[\omega]$ and $\delta \hat{Y}_{\delta\hat{a}_{{\rm{out}}}}[\omega]$. Importantly, these two quadratures are sufficient to construct a measurable correlation function containing the phonon noise contribution. This is in contrast to the heterodyne measurement technique used in Ref.~\cite{purdy_quantum_2017}. Instead, here, the low-frequency microwave signal allows the direct demodulation, simplifying measurement when compared to high-frequency optical measurements. The two quadratures can then be directly captured using a data acquisition system, following demodulation, without any additional post-processing. 

The phonon's noise contribution is included in the correlation spectrum via the component proportional to $\langle \{ \hat{Y}_{\delta \hat{a}} [\omega], \hat{Y}_{\delta \hat{a}} [\omega^\prime] \} \rangle$ and the temperature of the phonon mode can be determined by considering the ratio of two  conveniently chosen correlation spectra: one containing the above mentioned term and the other one a reference. For the former we notice that the phase-phase autocorrelation spectrum $S_{\frac{\pi}{2},\frac{\pi}{2}}[\omega]$, in the resolved sideband regime $\omega_{\rm{a,b,m}} \gg \kappa,\gamma_{\rm{m,b}}$, is given explicitly by
\begin{equation}
\begin{aligned}
    S_{\frac{\pi}{2}, \frac{\pi}{2} }[\omega] &= f_{\frac{\pi}{2}, \frac{\pi}{2} }[\omega] \vert(\chi_{\rm{b}}[\omega]-\chi_{\rm{b}}^*[-\omega])\vert^2 \times \\& \gamma_{\rm{b}}{\rm{coth}}\left(\frac{\hbar\omega}{2k_{\rm{B}} T}\right),
\label{eqn:SPi/2Pi/2}
\end{aligned}
\end{equation}
while the reference term is the amplitude-phase correlation spectrum $S_{0 ,\frac{\pi}{2} }[\omega]$ given by
\begin{equation}
    S_{0,\frac{\pi}{2}}[\omega] = i f_{0, \frac{\pi}{2} }[\omega] (\chi_{\rm{b}}[\omega]-\chi_{\rm{b}}^*[-\omega])(2n_{\rm{th}}+1).
\label{eqn:S0Pi/2}
\end{equation}
The frequency dependent coefficients $ f_{0, \frac{\pi}{2} }[\omega]$ and $f_{\frac{\pi}{2}, \frac{\pi}{2} }[\omega]$ contain information about the relevant coupling rates, as well as the photon and magnon susceptibilities. Moreover, at temperatures $T \ll \hbar \omega_{\rm{a,m}}/ k_{\rm{B}}$, $n_{\rm{th}} = 0$ and Eq.~\eqref{eqn:S0Pi/2} is temperature independent.

\begin{figure}[b]
\includegraphics[width=1.0\columnwidth]{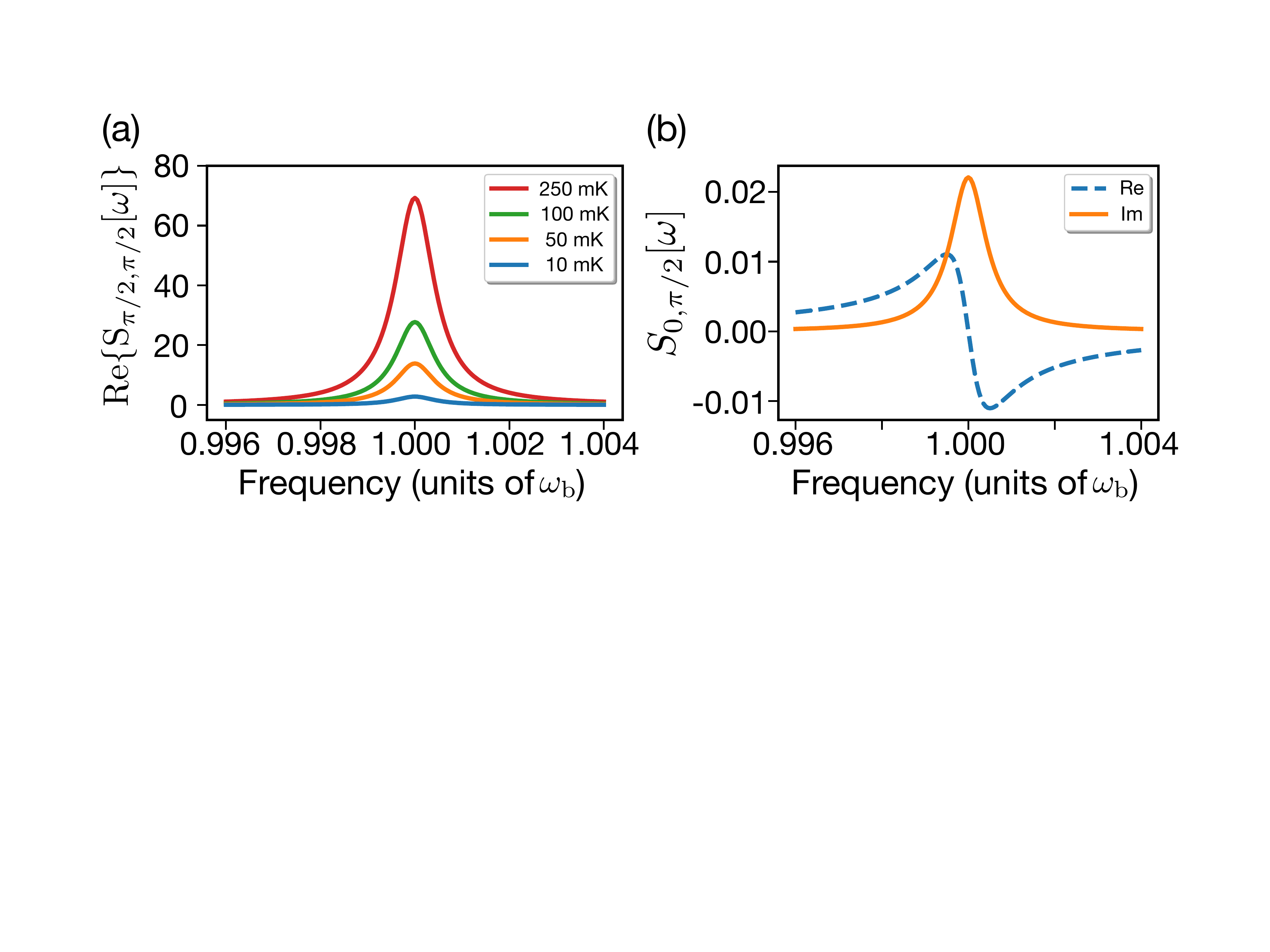}
\caption{a) Phase-phase autocorrelation spectrum for different temperatures. b) Amplitude-phase cross correlation spectrum, at 100 mK.  Each panel is plotted against frequency, normalized in units of $\omega_{\rm{b}}$.}
\label{Fig:02}
\end{figure}

Figure \ref{Fig:02} shows a calculated phase-phase autocorrelation spectrum (a) and an amplitude-phase correlation (b) as functions of the frequency. The maximum value of the phase-phase autocorrelation $S_{\frac{\pi}{2},\frac{\pi}{2}}[\omega]$ increases with the bath temperature $T$ and, similar to what was reported in Ref.~\cite{purdy_quantum_2017}, can be used as a thermometric measurement.

At low magnon and photon thermal occupancy, the terms related to photon and magnon shot noise in the function $f_{\frac{\pi}{2}, \frac{\pi}{2} }[\omega]$ can be ignored, and the phonon noise is the main component of the phase-phase autocorrelation \eqref{eqn:SPi/2Pi/2}. In this limit we have
\begin{equation}
    \frac{\operatorname{Re}\{S_{\frac{\pi}{2},\frac{\pi}{2}}[\omega]\}}{ \operatorname{Im}\{S_{0 ,\frac{\pi}{2} }[\omega]\}} = \frac{4{\rm{coth}}\left(\frac{\hbar\omega}{2k_{\rm{B}}T}\right)}{2n_{\rm{th}}+1},
\label{eqn:TempRatio}
\end{equation}
\noindent where the constant background contribution from $\operatorname{Re}\{S_{\frac{\pi}{2},\frac{\pi}{2}}[\omega]\}$ has been subtracted. This expression determines the temperature of the phonon mode via the measured correlation spectra and is independent of experimental parameters, such as coupling strengths and decay rates. We also note that the inclusion of all terms contained within $f_{\frac{\pi}{2}, \frac{\pi}{2} }[\omega]$ (see Eq.~\eqref{eqn:SPi/2Pi/2}) is consistent with Eq.~\eqref{eqn:TempRatio} within 0.1 mK for typical experimental parameters; see Appendix C for details.

Figure \ref{Fig:03} depicts the thermometric relation Eq.~\eqref{eqn:TempRatio} as a function of the phonon effective temperature, Fig.~3a, and for several values of the MW mode frequency, Fig.~3b. Although the relation defined in Eq.~\eqref{eqn:TempRatio} is unique for all temperatures, for $T > \hbar\omega_{\rm{a,m}}/k_{\rm{B}}$ the function is relatively flat. Therefore, the thermometric measurement will be most accurate at low-temperatures when the thermal photon/magnon occupation is less than one.

\begin{figure}[b]
\includegraphics[width=1.0\columnwidth]{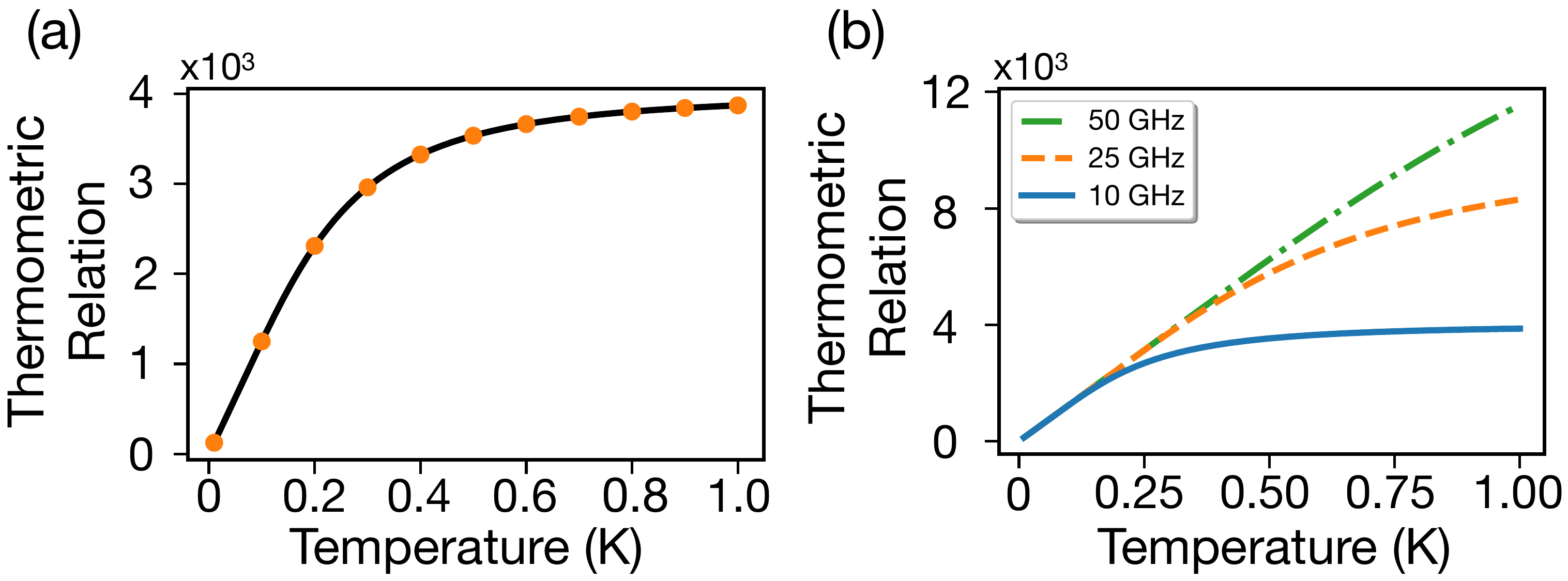}
\caption{a) Thermometric relationship as a function of bath temperature. The solid line represents the simplified analytical expression in Eq.~\eqref{eqn:TempRatio}, and the solid circles are numerically simulated values including all noise contributions within the phase-phase correlation function. For this curve $\omega_{\rm{b}} = 10$ MHz and $\omega_{\rm{a}}$ = $\omega_{\rm{m}}$ = 10 GHz. b) Thermometry relationship for different values of the microwave resonance frequency: $\omega_{\rm{m}} = 10, 25, 50$ GHz. In both plots $\omega=\omega_{\rm{b}}$, corresponding to the peak of $\textrm{Re}\{  S_{\frac{\pi}{2}, \frac{\pi}{2}} [\omega] \}$ and $\textrm{Im}\{  S_{0, \frac{\pi}{2}} [\omega] \}$.}
\label{Fig:03}
\end{figure}

It should be noted that heating due to photon absorption limits the effective lower temperature range of optical quantum-correlation thermometry to approximately 10 K \cite{purdy_quantum_2017,Meenehan_Pulsed_2015,hauer_two-level_2018,ramp_elimination_2019}. Therefore, the use of microwave photons, which cause minimal heating of the mechanical element due to their low energy allows this protocol to effectively be used at dilution temperatures below  $\sim$500 mK. Also, although a purely optomechanical quantum-correlation thermometry setup is possible within the microwave regime, it would suffer from similar loss of sensitivity at moderate temperatures due to the finite thermal occupation of the photon mode.

\subsection{Finite detuning effects and measurement considerations} 

\begin{figure}[b]
\includegraphics[width=0.73\columnwidth]{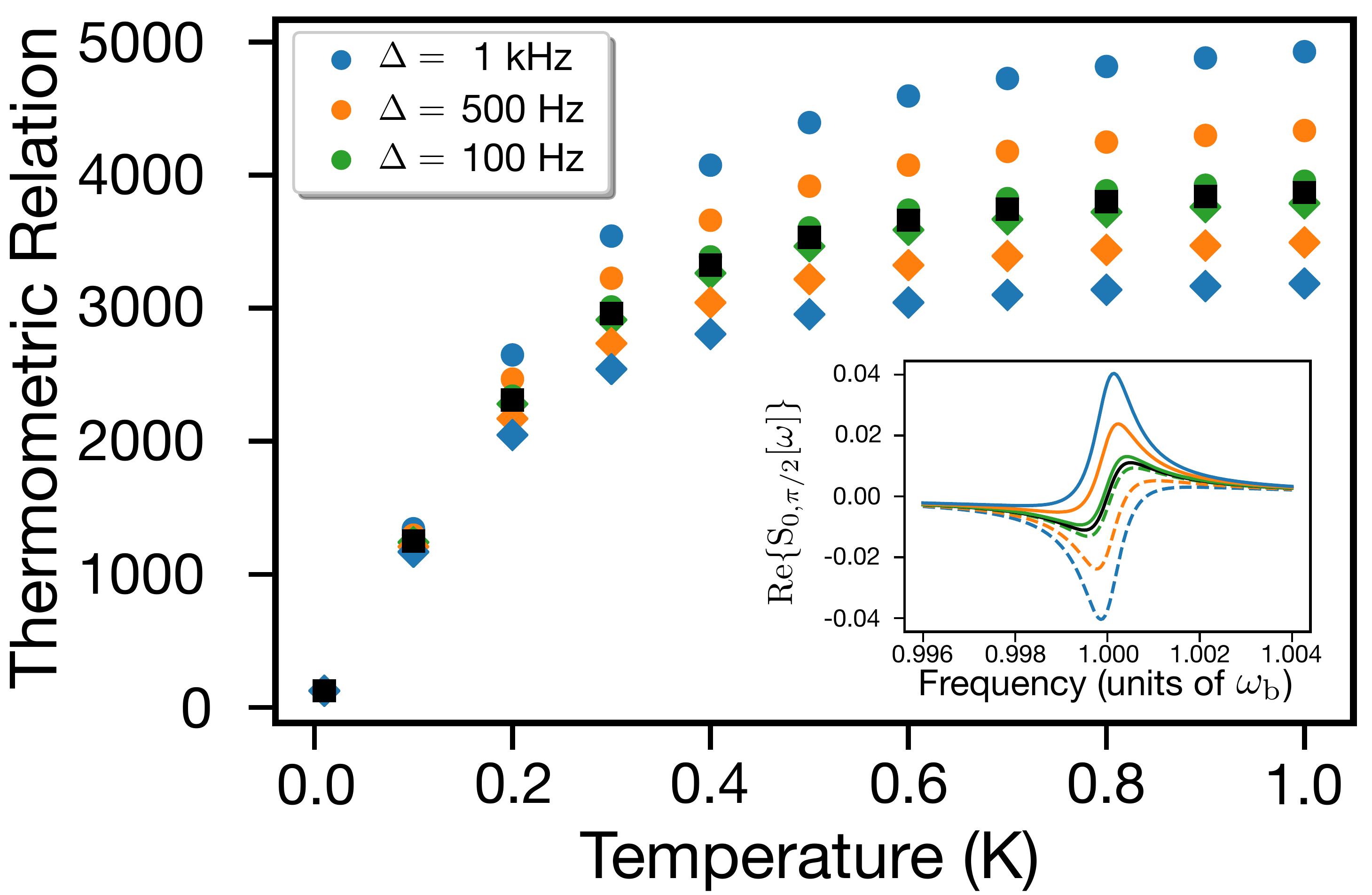}
\caption{Thermometric relation for different finite detunings of the MW drive. As the detuning increases $S_{0,\pi/2}[\omega]$ is contaminated with a contribution from the phonon thermal occupancy. Inset shows $\textrm{Re}\{S_{0,\pi/2}[\omega]\}$ for the same detunings. Diamond markers and dashed lines represent negative detunings.}
\label{Fig:06}
\end{figure}

We now address the effects of non-zero detuning of the MW drive in the proposed thermometric measurement. The thermometric relation Eq.~\eqref{eqn:TempRatio} was derived for $\omega_{\rm{d}} = \omega_{\rm{a}}$ (see Sec.~3). Finite values of the detuning introduce spurious effects. This is depicted in Fig.~\ref{Fig:06}, which shows the thermometric relation for different detunings. Experimentally these effects can be minimized by carefully varying the detuning and monitoring the real component of the amplitude-phase cross-correlation spectra. The peak-to-peak height of $S_{0,\pi/2}[\omega]$ directly depends on the value of the detuning. Therefore, minimizing the peak-to-peak height of the amplitude-phase cross-correlation spectra will minimize the thermometric relation error, as shown in the inset of Fig.~\ref{Fig:06}.

\begin{figure}[t]
\includegraphics[width=1.0\columnwidth]{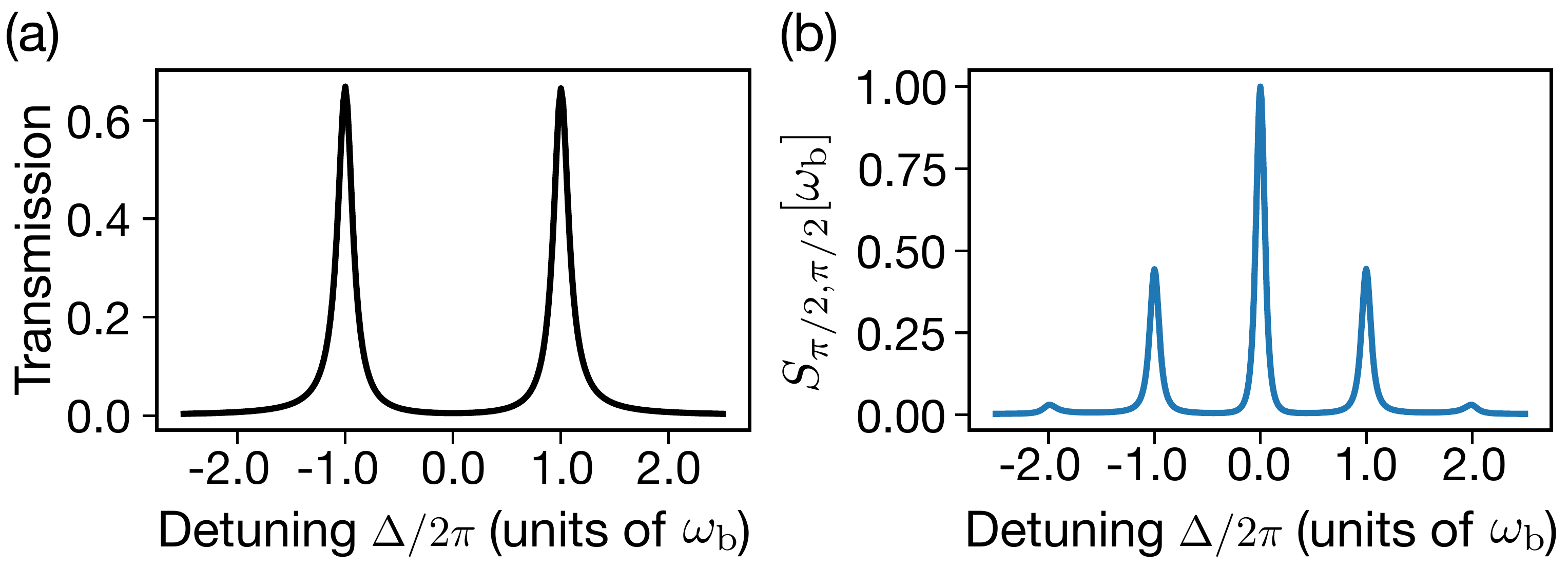}
\caption{a) Transmission spectrum of a strongly coupled magnon-photon system showing hybridized magnon-photon modes separated by $2g_{\rm{am}}$. b) Phase-phase correlation function for $\omega = \omega_{\rm{b}}$. The magnon-photon coupling rate was set to equal the phonon frequency, $g_{\rm{am}} = \omega_{\rm{b}}$, resulting in a peak in the correlation function at zero detuning, $\Delta_{\rm{a}} = 0$, corresponding to the measurement prescribed in the text.}
\label{Fig:04}
\end{figure}

\begin{figure}[b]
\includegraphics[width=0.78\columnwidth]{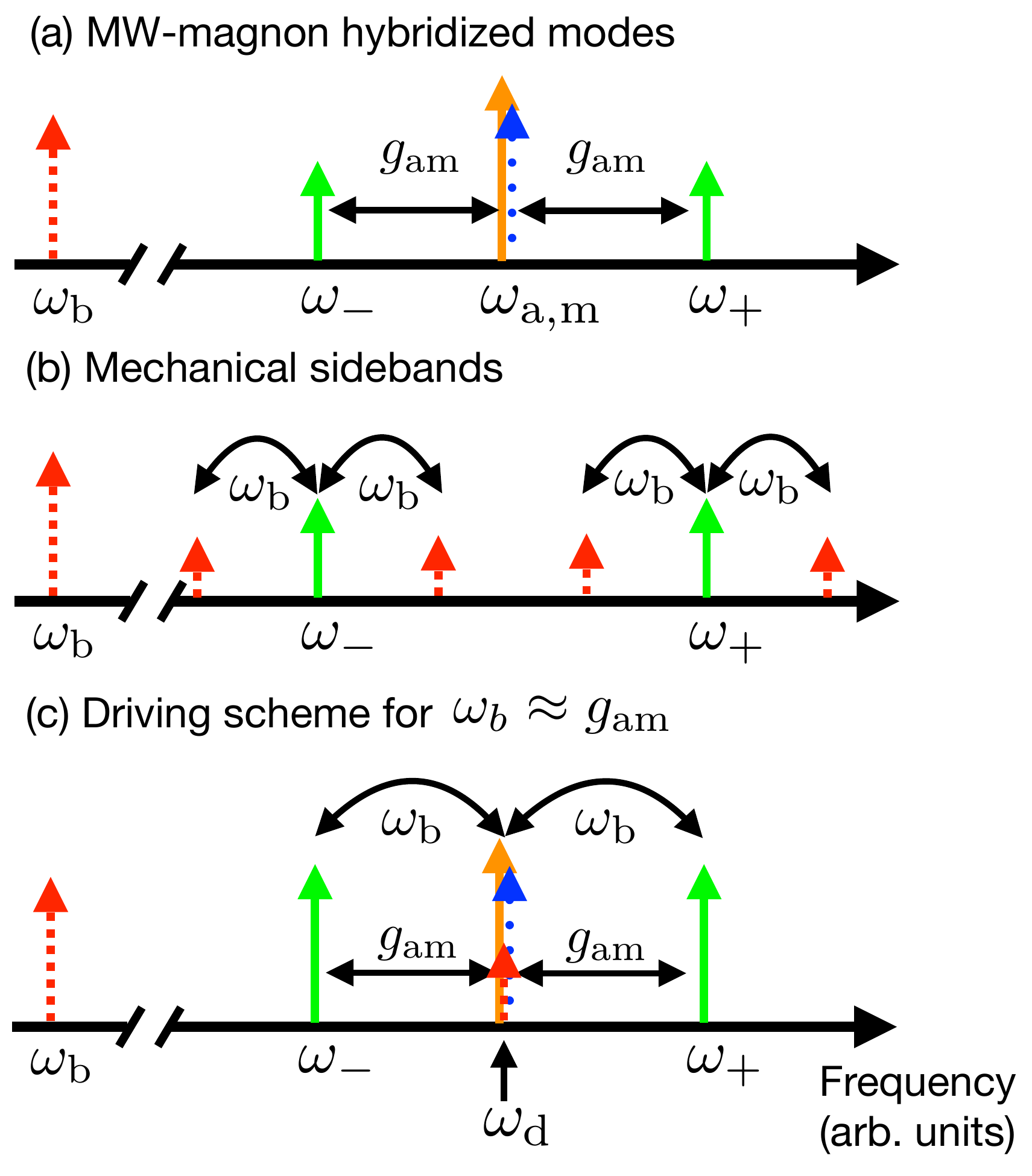}
\caption{Schematic illustration of the relevant frequencies in the system. a) The resonant magnon and photon modes (frequencies $\omega_{\rm{m}} \approx \omega_{\rm{a}}$) form two hybridized modes which, in the strong coupling regime considered ($g_{\rm{am}} \gg \kappa,\gamma_{\rm{m}}$), have frequencies $\omega_{\pm} \sim \omega_{\rm{a}(\rm{m})} \pm g_{\rm{am}}$. (b) Due to the interaction with the phonon mode, the hybrid modes have mechanical sidebands separated by $\omega_{\rm{b}}$ from their frequencies. (c) In our driving scheme we set $\omega_{\rm{b}} \approx g_{\rm{am}}$, which gives a cavity enhancement when the MW mode is pumped on resonance, despite initial expectations. This corresponds to pumping the mechanical sidebands of the hybridized MW-magnon modes.}
\label{Fig:FreqScheme}
\end{figure}

For the considered strong magnon-photon coupling, a drive tone tuned to $\Delta_{\rm{a}} = 0$ is far off-resonance. This is a consequence of the hybridization of the microwave and magnon modes forming two normal modes separated by $2g_{\rm{am}}$, as depicted in Fig.~\ref{Fig:04}. This leads to the wrong conclusion that it would be preferable to drive on resonance with the hybrid mode to allow an enhancement of the magnon-phonon coupling rate. Nevertheless, as discussed above, the thermometric relation in Eq.~\eqref{eqn:TempRatio} is precise for $\Delta_{\rm{a}} =0$. However, the signal-to-noise ratio can be improved by carefully tuning the magnon-photon coupling rate to match the frequency of the phonon mode, i.e. $g_{\rm{am}} = \omega_{\rm{b}}$. The coupling $g_{\rm{am}}$ depends on an overlap between the cavity mode and the magnetic element, and the aforementioned condition can be achieved by carefully positioning the magnetic element in the cavity \cite{Boventer_Complex_2018, Lachance_Hybrid_2019}. 
When this condition is satisfied, by pumping the cavity on resonance, one also pumps the mechanical sidebands of the hybrid modes as described in Fig.~\ref{Fig:FreqScheme}. The two sidebands constructively interfere producing an enhanced signal strength for the noise spectra, as shown in Fig.~\ref{Fig:04}b. Deviations from this ideal condition are shown in Appendix E.

\begin{figure}[h!]
\includegraphics[width=1.0\columnwidth]{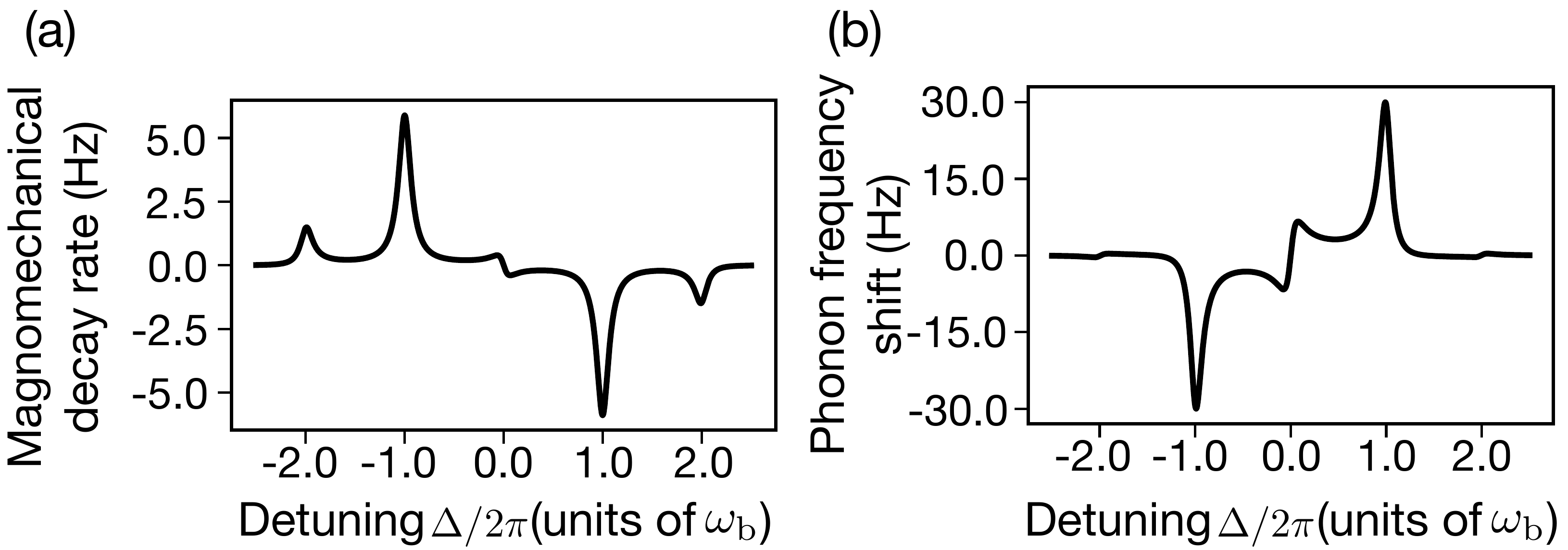}
\caption{a) Magnomechanical decay rate and b) phonon frequency shift (magnonic spring effect). For these plots $g_{\rm{am}} = \omega_{\rm{b}}$ and $G_{\rm{mb}} = 1$ kHz. The unique structure is due to interplay between multiple mechanical sidebands, as illustrated in Fig.~\ref{Fig:FreqScheme}. See Appendix E for additional plots.}
\label{Fig:05}
\end{figure}

Finally, accurate thermometry requires that the phonon mode is not affected by the microwave drive. This information is carried by the phonon self-energy term (see Appendix D) $\Sigma[\omega]$ \footnote{Where we adopt the self-energy notation as has been done in optomechanics due to the close analogy with how Dyson's equation modifies the bare Green's function due to interactions.}, given by
\begin{equation}
    \Sigma[\omega] = i\vert G_{\rm{mb}}\vert^2 ( \Xi[\omega] - \Xi^*[-\omega]),
\label{eqn:SelfEnergy}
\end{equation}
\noindent where $\Xi[\omega] = [\chi_{\rm{m}}^{-1}[\omega] + g_{\rm{am}}^2 \chi_{\rm{a}}[\omega]]$. In the weak magnon-phonon coupling limit, the mechanical frequency is shifted by $\delta\omega_{\rm{b}} = -\operatorname{Re}{\Sigma[\omega]}$ and the interaction induces an additional damping rate $\Gamma_{\rm{b}} = 2\operatorname{Im}{\Sigma[\omega]}$ which we refer to as the magnomechanical decay rate:
\begin{equation}
\begin{aligned}
    \tilde{\omega}_{\rm{b}} &= \omega_{\rm{b}} + \delta\omega_{\rm{b}},\\
    \tilde{\gamma}_{\rm{b}} &= \gamma_{\rm{b}} + \Gamma_{\rm{b}}.
\label{eqn:MagSpring}
\end{aligned}
\end{equation}
\noindent Fig.~\ref{Fig:05} shows the the frequency shift $\delta \omega_{\rm{b}}$ and the magnomechanical decay rate rate $\Gamma_{\rm{b}}$ of the phonon mode for the case of $g_{\rm{am}} = \omega_{\rm{b}}$. For $\Delta = 0$ it can be seen that both $\delta\omega_{\rm{b}}$ and $\Gamma_{\rm{b}}$ are zero. 

We can further consider the effect of the magnomechanical interaction for finite detunings, $\Delta_{\rm{a}} \ne 0$. As can be seen in Fig.~\ref{Fig:05} driving on the red(blue)-sideband, i.e. $\Delta_{\rm{a}} = n\omega_{\rm{b}}$, where $n$ is an integer, results in a positive (negative) additional damping rate, effectively cooling (heating) the mechanical mode. This effect is maximized for the first sideband and subsequent higher-order sidebands have a reduced additional damping rate. Additionally, we notice a shift in the mechanical resonance frequency as a result of the magnomechanical interaction. This frequency shift is analogous to the optical spring effect in optomechanics \cite{aspelmeyer_cavity_2014}, and we refer to it as the magnonic spring effect. The features depicted in Fig.~\ref{Fig:05} are highly dependent on the ratio $g_{\rm{am}}/\omega_{\rm{b}}$, as a result of the interplay between the mechanical sidebands and the hybridization of the photon-magnon modes. Here we have plotted these expressions for the experimentally relevant case $g_{\rm{am}}/\omega_{\rm{b}} = 1$; additional plots are shown in Appendix E.  

Finally, we point out that a colored-noise model, similar to the one used for the phonon mode, could also be adopted for the magnon mode. This would introduce detrimental effects for the thermometric relation, but such effects would only be relevant for low-quality factor magnons. A detailed discussion about such colored noise effects is presented in Ref.~\cite{giovannetti_phase_2001} in the context of optomechanics and thermomechanical motion, and a similar discussion should be valid for magnons.

\section{Conclusion}

In conclusion, we have proposed a thermometry method based on a magnon-phonon hybrid system in a microwave cavity. Through a correlation measurement scheme, we demonstrated how the backaction-induced mechanical spectrum can be used as a reference to calibrate the measured thermal correlation spectrum, allowing the determination of the temperature via the measured correlation spectra. The use of microwave photons reduces heating when compared to higher energy optical photons, making the method compatible with cryogenic temperatures. We discussed possible experimental sources of inaccuracies and showed that there is an upper-temperature limit for an accurate temperature reading, due to the effect of thermal photons and magnons.  All of the conditions considered here are compatible with current experimental capabilities and promise a straightforward platform for primary thermometry below 1 K, which we anticipate becoming widely adopted.

\begin{acknowledgments}
This work was supported by the University of Alberta, Faculty of Science; the Natural Sciences and Engineering Research Council, Canada (Grants No. RGPIN-04523-16, No. DAS-492947-16, and No. CREATE-495446-17); the Mitacs Globalink program; and the Max Planck Society.  
\end{acknowledgments}

\appendix

\section{Linearized quantum Langevin equations}

Starting with the Hamiltonain given by Eq.~\eqref{eq:Ham02}, we apply the unitary transformation $\hat{U} = {\rm{exp}}(i\omega_{\rm{d}}\hat{a}^{\dagger}\hat{a}+i\omega_{\rm{d}}\hat{m}^{\dagger}\hat{m})$ in order to remove the time-dependence from the driving term, $\hat{\mathcal{H}}^\prime=\hat{U}\hat{\mathcal{H}}\hat{U}^{\dagger} - i\hbar\hat{U}\partial\hat{U}^{\dagger}/\partial t$ \cite{aspelmeyer_cavity_2014} with
\begin{equation}
\begin{aligned}
\hat{\mathcal{H}}^\prime &= -\hbar\Delta_{\rm{a}}\hat{a}^{\dagger}\hat{a} + \hbar\omega_{\rm{b}}\hat{b}^{\dagger}\hat{b} - \hbar\Delta_{\rm{m}}\hat{m}^{\dagger}\hat{m} \\
&+\hbar g_{\rm{am}}(\hat{a}\hat{m}^{\dagger} + \hat{a}^{\dagger}\hat{m}) + \hbar g_{\rm{mb}}\hat{m}^{\dagger}\hat{m}(\hat{b}+\hat{b}^{\dagger}) \\
&+i \hbar\epsilon_{\rm{d}}\sqrt{\kappa_{\rm{p}}}(\hat{a}-\hat{a}^{\dagger}),
\end{aligned}
\label{eqn:TIHamApp}
\end{equation}
where $\Delta_{\rm{a}} = \omega_{\rm{d}} - \omega_{\rm{a}}$ and $\Delta_{\rm{m}} = \omega_{\rm{d}} - \omega_{\rm{m}}$ are the detunings between the drive and the cavity/magnon mode, $\kappa_{\rm{P}}$ is the coupling rate to the drive port and we have applied the rotating wave approximation. Moreover $\epsilon_{\rm{d}}=\sqrt{2 \kappa_{\rm{P}} \mathcal{P}/\hbar \omega_{\rm{d}}}$, with $\mathcal{P}$ the driving laser power.

Using the above Hamiltonian we derive the dynamics of any operator $\hat{\mathcal{O}}$  via the Heisenberg equation $-i\hbar \dot{\hat{\mathcal{O}}} = \left[ \hat{\mathcal{H}}^\prime, \hat{\mathcal{O}} \right]$, plus the addition of dissipation/fluctuation terms modelling the interaction with an environment. With those, ignoring the quantum fluctuations, we obtain the following semi-classical equations for the expectation values $\langle \hat{a} \rangle $, $\langle \hat{m} \rangle$ and $\langle \hat{b} \rangle$ 
\begin{equation}
\begin{aligned}
\langle\dot{ \hat{a}} \rangle &= \left(i\Delta_{\rm{a}} -\frac{\kappa}{2} \right) \langle \hat{a} \rangle  - i g_{\rm{am}}  \langle \hat{m} \rangle  - \epsilon_d \sqrt{\kappa_{\rm{P}}}, \\
\langle \dot{\hat{m}} \rangle &= \left( i \Delta_{\rm{m}} -\frac{\gamma_{\rm{m}}}{2} \right) \langle \hat{m} \rangle - i g_{\rm{am}} \langle \hat{a} \rangle \\&- i g_{\rm{mb}} \langle \hat{m}  \rangle( \langle \hat{b} \rangle  + \langle \hat{b}^\dagger \rangle), \\
\langle \dot{\hat{b}} \rangle&= \left(-i \omega_{\rm{b}} -\frac{\gamma_{\rm{b}}}{2}\right) \langle \hat{b} \rangle  - i g_{\rm{mb}} \vert \langle \hat{m} \rangle \vert^2.
\end{aligned}
\end{equation}
The classical steady state values $\langle \hat{a} \rangle$, $\langle \hat{m} \rangle$ and $\langle \hat{b} \rangle$ are then obtained by setting $\langle \dot{\hat{a}} \rangle = \langle \dot{\hat{b}} \rangle= \langle \dot{\hat{m}} \rangle= 0$. Additionnaly, we consider $g_{\rm{am}} \gg g_{\rm{mb}}$ such that
\begin{equation}
\begin{aligned}
\langle \hat{a} \rangle &= \frac{\left(i \Delta_{\rm{m}} - \gamma_{\rm{m}}/2 \right) \epsilon_d \sqrt{\kappa_{\rm{P}}}} {\left(i \Delta_{\rm{a}} - \kappa/2 \right)\left(i \Delta_{\rm{m}} - \gamma_{\rm{m}}/2 \right) +g_{\rm{am}}^2}, \\
\langle \hat{m} \rangle &= \frac{i g_{\rm{am}} \langle \hat{a} \rangle}{\left(i \Delta_{\rm{m}} - \gamma_{\rm{m}}/2 \right)}, \\
\langle \hat{b} \rangle &= -\frac{i g_{\rm{mb}} \vert \langle \hat{m} \rangle \vert^2}{i\omega_{\rm{b}} +\gamma_{\rm{b}}/2}.
\label{eqn:SteadyStateApp}
\end{aligned}
\end{equation}
Notice that at zero detuning $\Delta_{\rm{a}} = \Delta_{\rm{m}}=0$, since $\epsilon_{\rm{d}}$ is real; $\langle \hat{a} \rangle$ is real while $\langle \hat{m} \rangle$ is pure imaginary.

Next, we consider fluctuations around the steady state values \eqref{eqn:SteadyStateApp}: $\hat{a} = \langle \hat{a} \rangle + \delta \hat{a}$, $\hat{m} = \langle \hat{m} \rangle + \delta \hat{m}$ and $\hat{b} = \langle \hat{b} \rangle + \delta \hat{b}$. Neglecting high order terms in the fluctuations, we obtain the quadratic Hamiltonian

\begin{equation}
\begin{aligned}
    \hat{\mathcal{H}}_{\rm{Lin}} &= -\hbar\Delta_{\rm{a}}\delta\hat{a}^{\dagger}\delta\hat{a} + \hbar\omega_{b}\delta\hat{b}^{\dagger}\delta\hat{b} + -\hbar \tilde{\Delta}_{\rm{m}}\delta\hat{m}^{\dagger}\delta\hat{m} \\
    &+ \hbar g_{\rm{am}}(\delta\hat{a}\delta\hat{m}^{\dagger} + \delta\hat{a}^{\dagger}\delta\hat{m}) \\&+ \hbar ( G_{\rm{mb}} ^* \delta\hat{m} + G_{\rm{mb}} \delta\hat{m}^{\dagger})(\delta\hat{b}+\delta\hat{b}^{\dagger}),
    \label{eqn:LinHamApp}
\end{aligned}
\end{equation}
where, as defined in the main text,  $G_{\rm{mb}} = g_{\rm{mb}}\langle \hat{m} \rangle$ and  $\tilde{\Delta}_{\rm{m}}=\omega_d-\omega_{\rm{m}} - 2 \hbar g_{\rm{mb}} \mbox{Re}[\langle \hat{b} \rangle]$.

\section{Linear Langevin equations and the solutions in the frequency domain}

From the Hamiltonian, Eq.~\eqref{eqn:LinHam}, we obtain the linear coupled quantum Langevin equations
\begin{equation}
\begin{aligned}
    \delta\dot{\hat{a}} &= \left( i\Delta_{\rm{a}}  - \frac{\kappa}{2} \right)\delta\hat{a} -ig_{\rm{am}}\delta\hat{m} +\sqrt{\kappa_{\rm{P}}} \hat{\xi}_{\rm{P}}(t),\\
    \delta\dot{\hat{m}} &= \left( i\tilde{\Delta}_m - \frac{\gamma_{\rm{m}}}{2} \right)\delta\hat{m} -ig_{\rm{am}}\delta\hat{a} - iG_{\rm{mb}}(\delta\hat{b}+\delta\hat{b}^{\dagger}) \\ &+ \sqrt{\gamma_{\rm{m}}} \hat{\eta}(t),\\
    \delta\dot{\hat{b}} &= -\left(i\omega_{\rm{b}} + \frac{\gamma_{\rm{b}}}{2} \right)\delta\hat{b}  - i(G_{\rm{mb}}\delta\hat{m}^{\dagger} +G_{\rm{mb}}^*\delta\hat{m}) \\ &+\hat{\zeta}(t).
\label{eqn:EOMApp}
\end{aligned}
\end{equation}
These describe the evolution of the fluctuations, including the interaction with the environment via the noise operators $\hat{\xi}_{\rm{P}}(t), \hat{\eta}(t)$ and  $\hat{\zeta}(t)$  \cite{gardiner_quantum_2000}. In the time domain we have for $\hat{\beta}=\hat{\xi}_{{\rm P}},\hat{\eta}$:
\begin{equation}
\begin{aligned}
\langle\hat{\beta}(t)\hat{\beta}^{\dagger}(t^{\prime})\rangle	&=(n_{{\rm th}}+1)\delta(t-t^{\prime}), \\
\langle\hat{\beta}^{\dagger}(t)\hat{\beta}(t^{\prime})\rangle	&=n_{{\rm th}}\delta(t-t^{\prime}).
\end{aligned}
\end{equation}
These correlators describe the interaction of the photon/magnon modes with bosonic heat baths in the usual first Markovian approximation (see the main text). For the phonon mode we adopt the approach of Ref.~\cite{giovannetti_phase_2001} in which the environment effects are described in the framework of quantum Brownian motion. In this case, the correlator of the phonon noise reads
\begin{equation}
\begin{aligned}
\langle\hat{\zeta}(t)\hat{\zeta}^{\dagger}(t^{\prime})\rangle	&=\frac{1}{2\pi}\int d\omega e^{i\omega(t-t^{\prime})}\frac{\omega}{\omega_{\rm{b}}}(n[\omega]+1), \\
\langle\hat{\zeta}^{\dagger}(t)\hat{\zeta}(t^{\prime})\rangle	&= \frac{1}{2\pi}\int d\omega e^{i\omega(t-t^{\prime})}\frac{\omega}{\omega_{\rm{b}}}n[\omega],
\end{aligned}
\end{equation}
where $n[\omega]=[\exp(\hbar\omega/k_{{\rm B}}T)-1]^{-1}$ is the mean number of thermal phonons with frequency $\omega$ and temperature T.

We can write Eq.~\eqref{eqn:EOMApp} in the frequency domain by performing a Fourier transform  $\delta\hat{\mathcal{O}}[\omega] = \int_{-\infty}^{\infty} dt e^{ i\omega t} \delta\hat{\mathcal{O}}(t)$ and defining $\delta \hat{z}[\omega] = \delta \hat{b}[\omega]+\delta \hat{b}^\dagger[\omega]$:
\begin{equation}
\begin{aligned}
    \chi_{\rm{a}}^{-1}[\omega]\delta\hat{a}[\omega] &= -i g_{\rm{am}}\delta\hat{m}[\omega]+\sqrt{\kappa_{\rm{P}}} \hat{\xi}_{\rm{P}}[\omega], \\
    \chi_{\rm{m}}^{-1}[\omega]\delta\hat{m}[\omega] &= -i g_{\rm{am}}\delta\hat{a}[\omega] - iG_{\rm{mb}}\delta \hat{z}[\omega] +\sqrt{\gamma_{\rm{m}}} \hat{\eta}[\omega],\\
\delta\hat{z}[\omega] &= -i(\chi_{\rm{b}}[\omega]-\chi_{\rm{b}}^*[-\omega])\times\\
    &\Big[ G_{\rm{mb}}\delta\hat{m}^{\dagger}[-\omega]+G_{\rm{mb}}^*\delta\hat{m}[\omega] + \delta\hat{F}_{\rm{th}}[\omega]\Big],
\end{aligned}
\label{eqn:FreqEOMApp}
\end{equation}
where  $\chi_{\rm{a}}(\omega) = [-i(\Delta_{\rm{a}} + \omega) + \kappa/2]^{-1}$, $\chi_{\rm{m}}(\omega) = [-i(\tilde{\Delta}_{\rm{m}} + \omega) + \gamma_{\rm{m}}/2]^{-1}$ and $\chi_{\rm{b}}(\omega) = [i(\omega_{\rm{b}} - \omega) + \gamma_{\rm{b}}/2]^{-1}$ are the susceptibilities. The correlators of the noise operators $\hat{\xi}_{\rm{P}}[\omega]$ and $\hat{\eta}[\omega]$ in the frequency domain are given by Eq.~\eqref{eqn:PhotonNoise}, while the phonon noise $\delta \hat{F}_{\rm{th}}$ has correlation given by
\begin{equation}
\langle\delta\hat{F}_{{\rm {th}}}[\omega]\delta\hat{F}_{{\rm {th}}}[\omega^{\prime}]\rangle=2\pi\gamma_{{\rm b}} \frac{\omega}{\omega_{\rm{b}}} {\rm coth}\left(\frac{\hbar\omega}{2k_{{\rm B}}T}\right)\delta(\omega+\omega^{\prime}).
\label{eqn:SpectrumApp}
\end{equation}

By solving the linear system we obtain following solution for $\delta \hat{z}[\omega]$ in terms of only noise operators
\begin{widetext}
\begin{equation}
\begin{aligned}
    \delta\hat{z}[\omega]  &= \big[1+(\chi_{\rm{b}}[\omega]-\chi_{\rm{b}}^*[-\omega]) \vert G_{\rm{mb}} \vert ^2(\Xi[\omega]-\Xi^*[-\omega])\big]^{-1} \\& \times\Big[-i(\chi_{\rm{b}}[\omega]-\chi_{\rm{b}}^*[-\omega])(G_{\rm{mb}}\Xi^*[-\omega]\sqrt{\gamma_{\rm{m}}}\hat{\eta}^{\dagger}[-\omega]+G_{\rm{mb}}^*\Xi[\omega]\sqrt{\gamma_{\rm{m}}}\hat{\eta}[\omega])\\&+g_{\rm{am}}(\chi_{\rm{b}}[\omega]-\chi_{\rm{b}}^*[-\omega])\big[G_{\rm{mb}}\chi_{\rm{m}}^*[-\omega]\Lambda_{\rm{am}}^*[-\omega] \sqrt{\kappa_{\rm{P}}} \hat{\xi}^{\dagger}_{\rm{P}}[-\omega]\\&- G_{\rm{mb}}^*\chi_{\rm{m}}[\omega]\Lambda_{\rm{am}}[\omega] \sqrt{\kappa_{\rm{P}}} \hat{\xi}_{\rm{P}}[\omega] \big] -i(\chi_{\rm{b}}[\omega]-\chi_{\rm{b}}^*[-\omega])\delta\hat{F}_{\rm{th}}[\omega]\Big],
\label{eqn:deltaz}
\end{aligned}
\end{equation}
\end{widetext}
with $\Lambda_{\rm{am}}[\omega]=\big[\chi_{\rm{a}}^{-1}[\omega] + g_{\rm{am}}^2 \chi_{\rm{m}}[\omega] \big]^{-1}$ and $\,\, \Xi [\omega] = \Lambda_{\rm{am}}[\omega]\chi_{\rm{m}}[\omega]/\chi_{\rm{a}}[\omega]$.

Up to this point we have assumed an arbitrary detuning, however from now on we consider that the drive is on resonance with the cavity, such that the detuning is zero $\Delta_{\rm{a}} = \Delta_{\rm{m}} \equiv \Delta = 0$. We will also assume that the mechanical motion is within the sideband-resolved regime: $\omega_{\rm{b}} \gg \gamma_{\rm{m}},\gamma_{\rm{b}},\kappa$ \cite{zhang_cavity_2016}. Using these simplifying assumptions and the fact that, at zero detuning, the magnon steady state amplitude is purely imaginary, we have $G_{\rm{mb}}=i \vert G_{\rm{mb}} \vert$ and  $\chi_{a}[\omega] = \chi^*_a[-\omega],\chi_{m}[\omega] = \chi^*_m[-\omega],\Lambda_{\rm{am}}[\omega] = \Lambda_{\rm{am}}^*[-\omega]$ and $\Xi[\omega] = \Xi^*[-\omega]$. Notice however that $\chi_{b}[\omega] \ne \chi^*_b[-\omega]$. These simplifications allow $\delta \hat{z}[\omega]$ to be written in the form, 
\begin{equation}
\begin{aligned}
    \delta\hat{z}[\omega] &= i \vert G_{\rm{am}} \vert f_{\rm{m}}[\omega]\sqrt{\gamma_{\rm{m}}}(\hat{\eta}[\omega]-\hat{\eta}^{\dagger}[-\omega]) \\&+  \vert G_{am} \vert f_{\rm{a}}[\omega]\sqrt{\kappa_{\rm{P}}}(\hat{\xi}_{\rm{P}}[\omega]+\hat{\xi}_{\rm{P}}^{\dagger}[-\omega])\\&-i(\chi_{\rm{b}}[\omega]-\chi_{\rm{b}}^*[-\omega]) \delta \hat{F}_{\rm{th}}[\omega].
\label{eqn:deltazSimpApp}
\end{aligned}
\end{equation}
In this expression we have defined $ f_{\rm{m}}[\omega] = i\Xi[\omega](\chi_{\rm{b}}[\omega]-\chi_{\rm{b}}^*[-\omega])$ and $f_{\rm{a}}[\omega] = ig_{\rm{am}}\chi_{\rm{m}}[\omega]\Lambda_{\rm{am}}[\omega](\chi_{\rm{b}}[\omega]-\chi_{\rm{b}}^*[-\omega])$, such that $f_{a,m}[\omega] = f_{a,m}^*[-\omega]$.

The microwave cavity field operator can be written in terms of noise operators by inserting Eq.~\eqref{eqn:deltazSimpApp} into Eq.~\eqref{eqn:Deltaa} to get
\begin{equation}
\begin{aligned}
    \delta\hat{a}[\omega]  &= \Lambda_{\rm{am}}[\omega] g_{\rm{am}}\chi_{\rm{m}}[\omega] \Big[ -i\sqrt{\gamma_{\rm{m}}}\hat{\eta}[\omega] \\ &+\vert G_{\rm{mb}} \vert ^2 f_{\rm{m}}[\omega] \sqrt{\gamma_{\rm{m}}}\big(\hat{\eta}[\omega]  -\hat{\eta}^{\dagger}[-\omega]\big)\\&-i \sqrt{\kappa_{\rm{P}}} \vert G_{\rm{mb}} \vert^2 f_{\rm{a}}[\omega]\big(\hat{\xi}_{\rm{P}}[\omega]+\hat{\xi}_{\rm{P}}^{\dagger}[-\omega]\big) \\ &- (\chi_{\rm{b}}[\omega]-\chi_{\rm{b}}^*[-\omega])  \vert G_{\rm{mb}} \vert \delta \hat{F}_{\rm{th}}[\omega]\Big]\\&+\sqrt{\kappa_{\rm{P}}}\Lambda_{\rm{am}}[\omega]\hat{\xi}_{\rm{P}}[\omega].
\label{eqn:deltaaApp}
\end{aligned}
\end{equation}
\noindent Using Eq.~\eqref{eqn:deltaaApp} we can proceed with calculating the optical quadratures in Appendix C. 

\section{Calculation of the Quadrature Correlations}

The calculation of the phase-phase autocorrelation function Eq.~\eqref{eqn:SPi/2Pi/2} and the amplitude-phase cross-correlation function Eq.~\eqref{eqn:S0Pi/2} requires the evaluation of Eq.~\eqref{eqn:NoiseSpec}, the symmetrized correlation spectrum. For this we need to evaluate the two output quadratures, $\hat{X}_{\rm{out}}[\omega] =  \delta\hat{a}_{{\rm{out}}}[\omega] + \delta\hat{a}^\dagger_{{\rm{out}}}[\omega]$ and $\hat{Y}_{\rm{out}}[\omega]=-i \Big( \delta\hat{a}_{{\rm{out}}}[\omega] - \delta\hat{a}^\dagger_{{\rm{out}}}[\omega] \Big)$, given in terms of the output field
\begin{equation}
\delta\hat{a}_{{\rm{out}}}[\omega] = \hat{\xi}_{\rm{P}}[\omega] - \sqrt{\kappa_{\rm{P}}}\delta\hat{a}[\omega].
\label{eq:OutApp} 
\end{equation}
From Eq.~\eqref{eqn:deltaaApp} and the input-output relation Eq.~\eqref{eq:OutApp} we can construct the required quadratures and calculate the noise spectra via the expectation values of products of quadratures. 

For deriving the thermometric relation we need to consider the correlation spectra, Eq.~\eqref{eqn:SPi/2Pi/2} and Eq.~\eqref{eqn:S0Pi/2}, which are given in terms of the following expectation values
\begin{equation}
\begin{aligned}
    \langle  \hat{Y}_{\rm{out}} [\omega] \hat{Y}_{\rm{out}} [\omega'] \rangle &= -\sqrt{\kappa_P} \langle  \hat{Y}_{\rm{in}} [\omega]\hat{Y}_{\delta\hat{a}}[\omega^\prime] \rangle\\& - \sqrt{\kappa_P} \langle \hat{Y}_{\delta\hat{a}}[\omega] \hat{Y}_{\rm{in}} [\omega^\prime] \rangle\\&+\kappa_P\langle \hat{Y}_{\delta\hat{a}}[\omega]\hat{Y}_{\delta\hat{a}}[\omega^\prime] \rangle, \\ 
    \langle  \hat{X}_{\rm{out}} [\omega] \hat{Y}_{\rm{out}} [\omega^\prime] \rangle &= -\sqrt{\kappa_P}  \langle  \hat{X}_{\rm{in}} [\omega]\hat{Y}_{\delta\hat{a}}[\omega^\prime] \rangle \\ &-\sqrt{\kappa_P}  \langle \hat{X}_{\delta\hat{a}}[\omega] \hat{Y}_{\rm{in}} [\omega^\prime] \rangle )\\&+\kappa_P\langle \hat{X}_{\delta\hat{a}}[\omega]\hat{Y}_{\delta\hat{a}}[\omega^\prime] \rangle,
\label{eqn:ExpOutApp}
\end{aligned}
\end{equation}
where $\hat{Y}_{\rm{in}} = -i(\hat\xi_P - \hat\xi^{\dagger}_P)$ and $ \hat{X}_{\rm{in}}  = (\hat\xi_P + \hat\xi^{\dagger}_P)$ are the input noise quadratures for the measurement port. The expectation values contained within Eq.~\eqref{eqn:ExpOutApp} are given in terms of expectation values of noise quadratures by 

\begin{widetext}
\begin{equation}
\begin{aligned}
    \langle  \hat{Y}_{\rm{in}} [\omega]\hat{Y}_{\delta\hat{a}}[\omega^\prime ] \rangle & = 2 \pi \sqrt{\kappa_P} \Lambda_{\rm{am}}[\omega^\prime] \delta(\omega+\omega^\prime) \Big( (2n_{\rm{th}}+1) +2 i g_{\rm{am}} \vert G_{\rm{mb}} \vert ^2 \chi_{\rm{m}}[\omega^\prime]f_{\rm{a}}[\omega^\prime ] \Big), \\
   \langle \hat{X}_{\rm{in}} [\omega]\hat{Y}_{\hat{a}}[\omega^\prime ]\rangle &= 2 i \pi \sqrt{\kappa_P}\Lambda_{\rm{am}}[\omega^\prime]\delta(\omega+\omega^\prime) \Big(2i g_{\rm{am}} \vert G_{\rm{mb}} \vert ^2 \chi_{\rm{m}}[\omega^\prime]f_{\rm{a}}[\omega^\prime](2n_{\rm{th}}+1) +1\Big), \\
    \langle \hat{Y}_{\delta\hat{a}}[\omega] \hat{Y}_{\delta\hat{a}}[\omega^\prime] \rangle &= 2 g_{\rm{am}}^2 \vert G_{\rm{mb}} \vert^2 \Lambda_{\rm{am}}[\omega]\Lambda_{\rm{am}}[\omega^\prime]\chi_{\rm{m}}[\omega]\chi_{\rm{m}}[\omega^\prime] \Big(  2 \gamma_{\rm{m}} \vert G_{\rm{mb}} \vert^2 f_{\rm{m}}[\omega]f_{\rm{m}}[\omega^\prime] \langle \delta \hat{Y}_{\hat{\eta}}[\omega]\delta\hat{Y}_{\hat{\eta}}[\omega^\prime] \rangle \\ &-  \gamma_{\rm{m}} f_{\rm{m}}[\omega]\langle \delta\hat{Y}_{\hat{\eta}}[\omega]\delta\hat{X}_{\hat{\eta}}[\omega^\prime]\rangle - \gamma_{\rm{m}} f_{\rm{m}}[\omega^\prime] \langle \delta \hat{X}_{\hat{\eta}}[\omega]\delta\hat{Y}_{\hat{\eta}}[\omega^\prime]\rangle  + 2 \kappa_{\rm{P}} f_{\rm{a}}[\omega]f_{\rm{a}}[\omega^\prime ]\langle\delta\hat{X}_{\hat{\xi}_{\rm{P}}}[\omega]\delta\hat{X}_{\hat{\xi}_{\rm{P}}}[\omega^\prime ]\rangle \Big) \\
    &+g_{\rm{am}} \Lambda_{\rm{am}}[\omega]\Lambda_{\rm{am}}[\omega^\prime] \Big( \gamma_{\rm{m}} g_{\rm{am}} \chi_{\rm{m}}[\omega]\chi_{\rm{m}}[\omega^\prime]\langle\delta\hat{X}_{\hat{\eta}}[\omega]\delta\hat{X}_{\hat{\eta}}[\omega^\prime]\rangle - 2 \kappa_{\rm{P}}  \vert G_{\rm{mb}} \vert ^2 \chi_{\rm{m}}[\omega]f_{\rm{a}}[\omega]\langle\delta\hat{X}_{\hat{\xi}_{\rm{P}} }[\omega]\delta\hat{Y}_{\hat{\xi}_{\rm{P}} }[\omega^\prime]\rangle \\ &- 2 \kappa_{\rm{P}}   \vert G_{\rm{mb}} \vert ^2 \chi_{\rm{m}}[\omega^\prime ]f_{\rm{a}}[\omega^\prime ]\langle\delta\hat{Y}_{\hat{\xi}_{\rm{P}}}[\omega]\delta\hat{X}_{\hat{\xi}_{\rm{P}}}[\omega^\prime]\rangle \Big) + \kappa_{\rm{P}} \Lambda_{\rm{am}}[\omega]\Lambda_{\rm{am}}[\omega^\prime ]\langle \delta\hat{Y}_{\hat{\xi}_{\rm{P}} }[\omega]\delta\hat{Y}_{\hat{\xi}_{\rm{P}} }[\omega^\prime]\rangle \\ &+ 4 g_{\rm{am}}^2 \vert G_{\rm{mb}} \vert ^2(\chi_{\rm{b}}[\omega]-\chi_{\rm{b}}^*[-\omega])(\chi_{\rm{b}}[\omega^\prime]-\chi_{\rm{b}}^*[-\omega^\prime]) \Lambda_{\rm{am}}[\omega]\Lambda_{\rm{am}}[\omega^\prime]\chi_{\rm{m}}[\omega]\chi_{\rm{m}}[\omega^\prime]\langle \delta\hat{F}_{\rm{th}}[\omega]\delta\hat{F}_{\rm{th}}[\omega^\prime] \rangle,\\ 
    \langle\hat{X}_{\delta\hat{a}}[\omega]\hat{Y}_{\delta\hat{a}}[\omega^\prime ]\rangle&= \gamma_{\rm{m}} g_{\rm{am}}^2 \Lambda_{\rm{am}}[\omega]\Lambda_{\rm{am}}[\omega^\prime ]\chi_{\rm{m}}[\omega]\chi_{\rm{m}}[\omega^\prime ] \Big( - \langle\delta\hat{Y}_{\hat{\eta}}[\omega]\delta\hat{X}_{\hat{\eta}}[\omega^\prime ]\rangle + 2 \vert G_{\rm{mb}} \vert ^2 f_{\rm{m}}[\omega^\prime ] \langle\delta\hat{Y}_{\hat{\eta}}[\omega]\delta\hat{Y}_{\hat{\eta}}[\omega^\prime ]\rangle \Big) \\ &+ \kappa_{\rm{P}} \Lambda_{\rm{am}}[\omega]\Lambda_{\rm{am}}[\omega^\prime ] \Big( \langle\delta\hat{X}_{\hat{\xi}_{\rm{P}}}[\omega]\delta\hat{Y}_{\hat{\xi}_{\rm{P}}}[\omega^\prime ]\rangle -2g_{\rm{am}} \vert G_{\rm{mb}} \vert ^2 \chi_{\rm{m}}[\omega^\prime ]f_{\rm{a}}[\omega^\prime ]\langle\delta\hat{X}_{\hat{\xi}_{\rm{P}}}[\omega]\delta\hat{X}_{\hat{\xi}_{\rm{P}}}[\omega^\prime ]\rangle   \Big)
\label{eqn:CorFunApp}
\end{aligned}
\end{equation}
\end{widetext}

\noindent The expectation values for the phonon and magnon noise quadratures can be calculated using Eq.~\eqref{eqn:PhotonNoise} and are (for $\hat{\beta} = \hat{\xi}_{\rm{P}}, \hat{\eta}$), 
\begin{equation}
\begin{aligned}
    \langle \delta\hat{X}_{\hat{\beta}}[\omega]\delta\hat{X}_{\hat{\beta}}[\omega^\prime ] \rangle &= \langle \delta\hat{Y}_{\hat{\beta}}[\omega]\delta\hat{Y}_{\hat{\beta}}[\omega^\prime ] \rangle\\ &= 2 \pi (2n_{\rm{th}} +1) \delta(\omega+\omega^\prime ),\\
     \langle \delta\hat{X}_{\hat{\beta}}[\omega]\delta\hat{Y}_{\hat{\beta}}[\omega^\prime ] \rangle &= - \langle \delta\hat{Y}_{\hat{\beta}}[\omega]\delta\hat{X}_{\hat{\beta}}[\omega^\prime ] \rangle \\ &= i 2 \pi \delta(\omega+\omega^\prime ),
\label{eqn:NoiseCorrApp}
\end{aligned}
\end{equation}
while the phonon noise correlator is given by Eq.~\eqref{eqn:SpectrumApp}

For all plots in the main text we used the full expressions given by Eq.~\eqref{eqn:CorFunApp}. However, a simplified relation can be obtained by ignoring all terms related to the photon and magnon shot noises within the expression for $\langle \hat{Y}_{\delta\hat{a}}[\omega] \hat{Y}_{\delta\hat{a}}[\omega^\prime ] \rangle$. Since we are considering the experimentally relevant resolved-sideband regime, all the terms, besides the phonon noise correlation, contained within $\langle \hat{Y}_{\delta\hat{a}}[\omega] \hat{Y}_{\delta\hat{a}}[\omega^\prime ] \rangle$ are sharply peaked around $\omega = 0$ and for $\omega$ around $\omega_{\rm{b}}$ the only relevant contribution will be the phonon noise term. In this case
\begin{equation}
\begin{aligned}
    \langle \hat{Y}_{\delta\hat{a}}[\omega] \hat{Y}_{\delta\hat{a}}[\omega^\prime ] \rangle &\approx 4g_{\rm{am}}^2 \vert G_{\rm{mb}} \vert ^2\\&\vert(\chi_{\rm{b}}[\omega]-\chi_{\rm{b}}^*[-\omega])\vert^2\\&\Lambda_{\rm{am}}[\omega]\Lambda_{\rm{am}}[\omega^\prime ]\chi_{\rm{m}}[\omega]\chi_{\rm{m}}[\omega^\prime ]\\&\langle \delta\hat{F}_{\rm{th}}[\omega]\delta\hat{F}_{\rm{th}}[\omega^\prime ] \rangle.
\end{aligned}
\end{equation}

\begin{figure}[b]
\includegraphics[width=0.55\columnwidth]{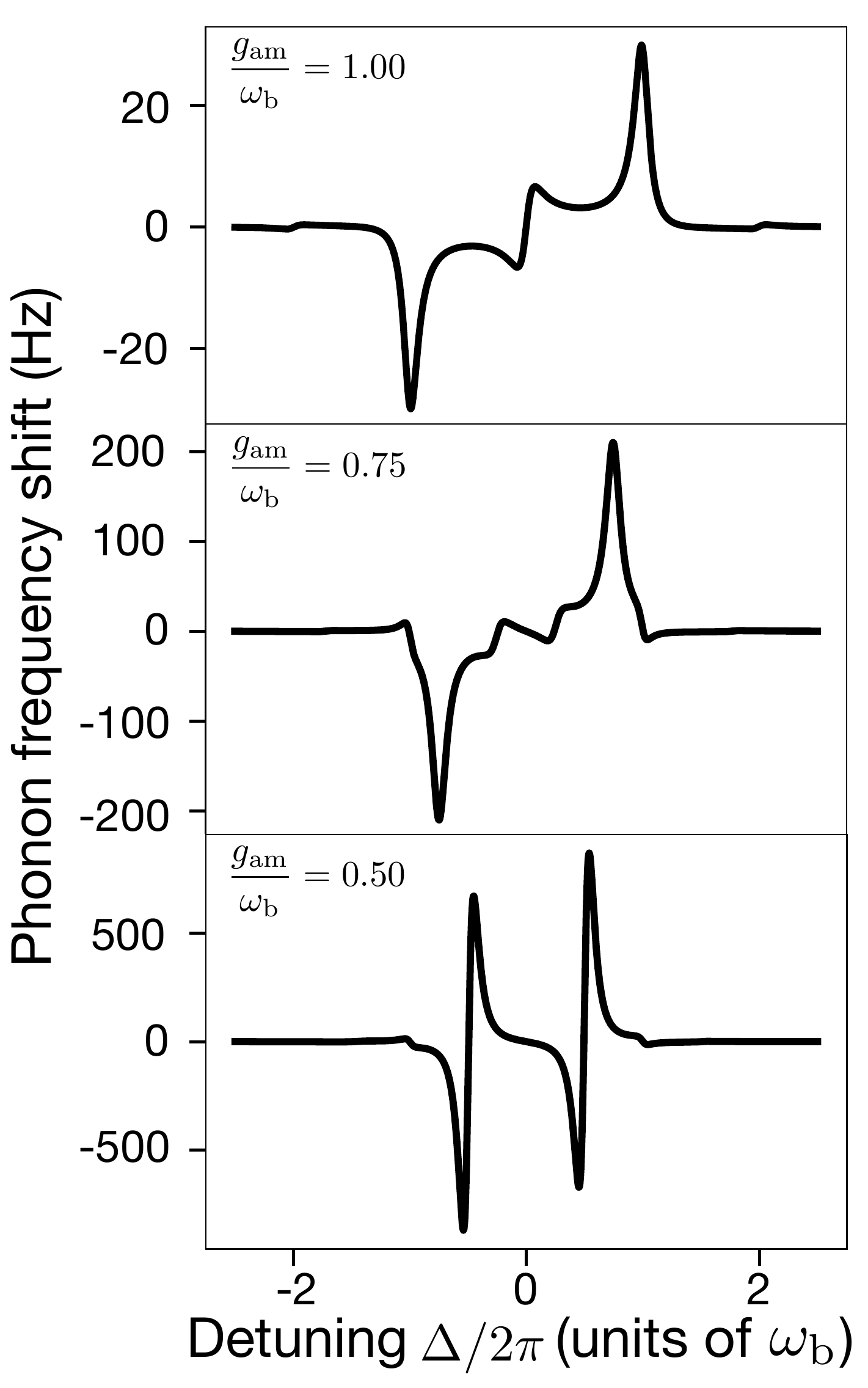}
\caption{Phonon frequency shift as function of the drive-cavity detuning for $\omega=\omega_{\rm{b}}$ and several values of $g_{\rm{am}}/\omega_{\rm{b}}$.
}\label{Fig:App01}
\end{figure}

Using these definitions we can construct the symmetrized expectation values $\langle \{ \mathcal{\hat{A}}[\omega],\mathcal{\hat{B}}[\omega^\prime] \} \rangle = (\langle \mathcal{\hat{A}}[\omega]\mathcal{\hat{B}}[\omega^\prime] \rangle +\langle \mathcal{\hat{B}}[\omega^\prime]\mathcal{\hat{A}}[\omega] \rangle)/2$. It is necessary to use the symmetrized expectation value to compare with the classically accessible measurement currents. By performing the integration over frequency space and properly normalizing, as defined in Eq.~\eqref{eqn:NoiseSpec}, we arrive at the symmetrized noise spectra given by, 
\begin{equation}
    \begin{aligned}
        S_{\pi/2,\pi/2}[\omega] &= 2\pi \kappa_{\rm{P}} g_{\rm{am}}^2\vert G_{\rm{mb}}\vert^2 \vert \chi_{\rm{b}}[\omega]-\chi^*_b[-\omega]\vert^2 \times\\& \times \vert \Lambda_{\rm{am}}[\omega] \vert^2 \vert \chi_{\rm{m}}[\omega] \vert^2 \gamma_{\rm{b}} \frac{\omega}{\omega_{\rm{b}}} \rm{coth}\bigg(\frac{\hbar \omega}{2 \rm{k}_{\rm{B}}\rm{T}}\bigg)\\&-\pi\kappa_{\rm{P}} \Lambda_{\rm{am}}[\omega](2n_{\rm{th}}+1), \\ \\ S_{0,\pi/2}[\omega] &= \pi \kappa_{\rm{P}} g_{\rm{am}}^2 \vert G_{\rm{mb}} \vert^2 \Lambda_{\rm{am}}^2[-\omega]\chi_{\rm{m}}^2[-\omega]\\& i(\chi_{\rm{b}}[\omega]-\chi_{\rm{b}}[-\omega])(2n_{\rm{th}}+1)\\& [1+\Lambda_{\rm{am}}[\omega]\chi_{\rm{m}}[\omega]/\chi_{\rm{a}}[-\omega]-\Lambda_{\rm{am}}[\omega]].
        \label{eqn:PowSpecApp}
    \end{aligned}
\end{equation}

\noindent Here, the term $\pi \kappa_{\rm{P}} \Lambda_{\rm{am}}[\omega](2n_{\rm{th}}+1)$ is a constant offset that can be subtracted in post processing. Furthermore, $2\gamma_{\rm{b}}\vert \chi_{\rm{b}}[\omega]-\chi^*_b[-\omega]\vert^2 = i(\chi_{\rm{b}}[\omega]-\chi_{\rm{b}}[-\omega])$ and after some algebraic manipulation it can be shown that Eq.~\eqref{eqn:PowSpecApp} leads to Eq.~\eqref{eqn:TempRatio}.

\begin{figure}[t]
\includegraphics[width=0.55\columnwidth]{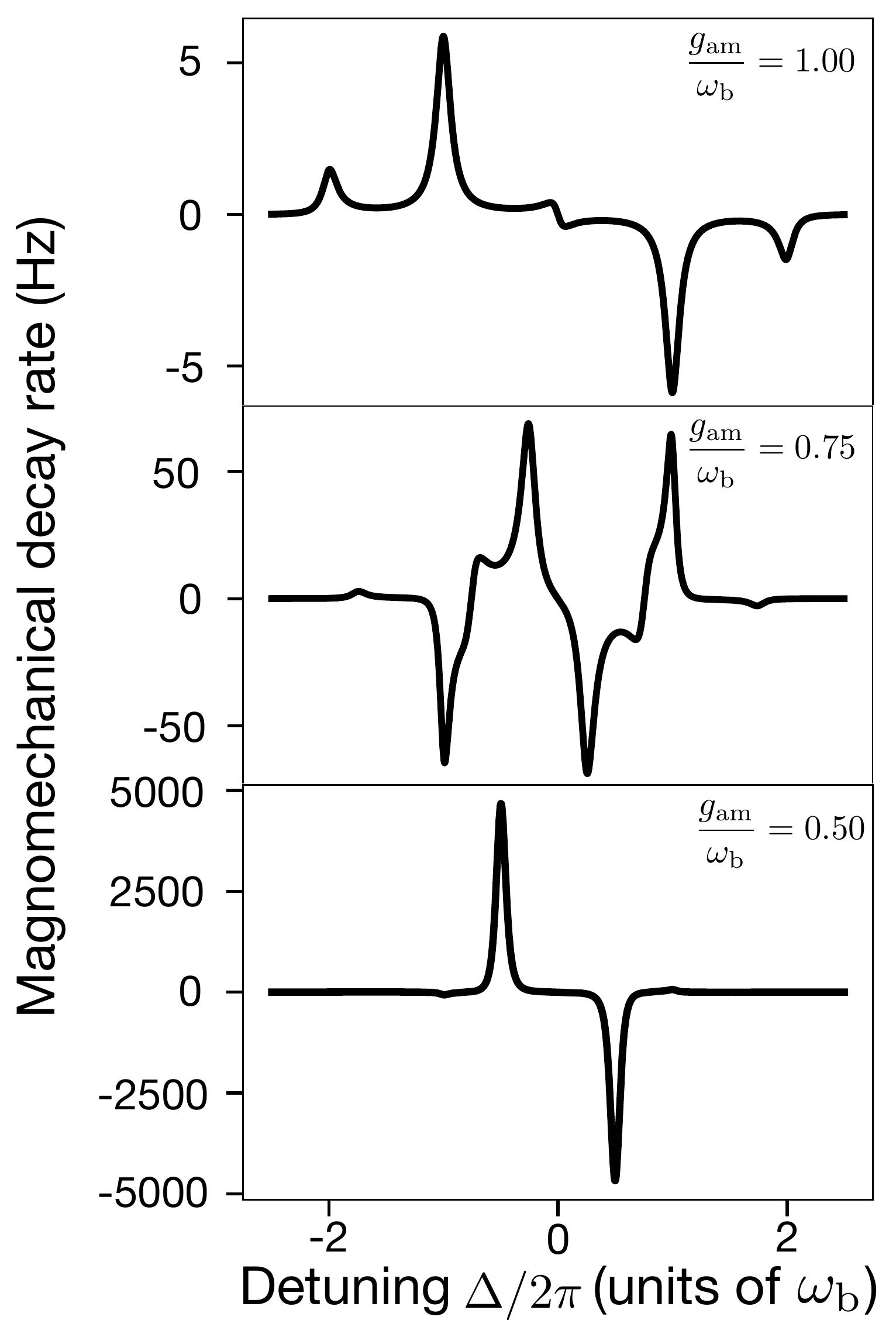}
\caption{Magnomechanical decay rate as function of the drive-cavity detuning for $\omega=\omega_{\rm{b}}$ and several values of $g_{\rm{am}}/\omega_{\rm{b}}$.
}\label{Fig:App02}
\end{figure}

\section{Phonon self-energy}

Starting from the equations of motion in the time domain Eq.~\eqref{eqn:EOMApp} we obtain the equation for the magnon mode in the frequency domain
\begin{equation}
    \chi_{\rm{b}}^{-1} \delta \hat{b}[\omega] = - i \left( G_{\rm{mb}} \delta \hat{m}^\dagger [\omega] + G^*_{\rm{mb}} \delta \hat{m} \right) + \hat{\zeta} [\omega].
    \label{D1}
\end{equation}
We then obtain a system of equations similar to Eq.~\eqref{eqn:FreqEOMApp} but with Eq.~\eqref{D1} and an equation for $\delta \hat{b}[-\omega]$ in place of $\delta \hat{z}[\omega]$. By solving this system we get:
\begin{equation}
\begin{aligned}
\chi_{\rm{b}}^{-1}[\omega] \delta \hat{b} [\omega] &= -\vert G_{\rm{mb}} \vert^2\left( \Xi[\omega] -\Xi^*[-\omega] \right)\bigg[1+  \\
&\frac{\vert G_{\rm{mb}} \vert^2(\Xi[\omega] -\Xi^*[-\omega])}{(\chi_{\rm{b}}^*)^{-1} - \vert G_{\rm{mb}} \vert^2(\Xi[\omega] -\Xi^*[-\omega])} \bigg] \delta \hat{b} [\omega] + \Upsilon,
\end{aligned}
\end{equation}
where the last term $\Upsilon$ represents all the noise terms. We rewrite this equation as
\begin{equation}
    \left(\chi_{\rm{b}}^{-1}[\omega] - i \Sigma[\omega] \right) \delta \hat{b} [\omega] = \Upsilon,
\end{equation}
where we identify the phonon self energy $\Sigma[\omega]$ as given by
\begin{equation}
\begin{aligned}
    \Sigma[\omega]&= i \vert G_{\rm{mb}} \vert^2\left( \Xi[\omega] -\Xi^*[-\omega] \right)\bigg[1+ \\ &\frac{\vert G_{\rm{mb}} \vert^2(\Xi[\omega] -\Xi^*[-\omega])}{(\chi_{\rm{b}}^*)^{-1} - \vert G_{\rm{mb}} \vert^2(\Xi[\omega] -\Xi^*[-\omega])} \bigg].
\end{aligned}
\end{equation}
Under the approximations used throughout this paper, but without restricting the drive-detuning to zero, the self energy is given by $\Sigma[\omega] \approx i  \vert G_{\rm{mb}} \vert^2\left( \Xi[\omega] -\Xi^*[-\omega] \right)$, as in Eq.~\eqref{eqn:SelfEnergy}.

\section{Measurement effects of photon-magnon coupling rate}

\begin{figure}[b]
\includegraphics[width=0.55\columnwidth]{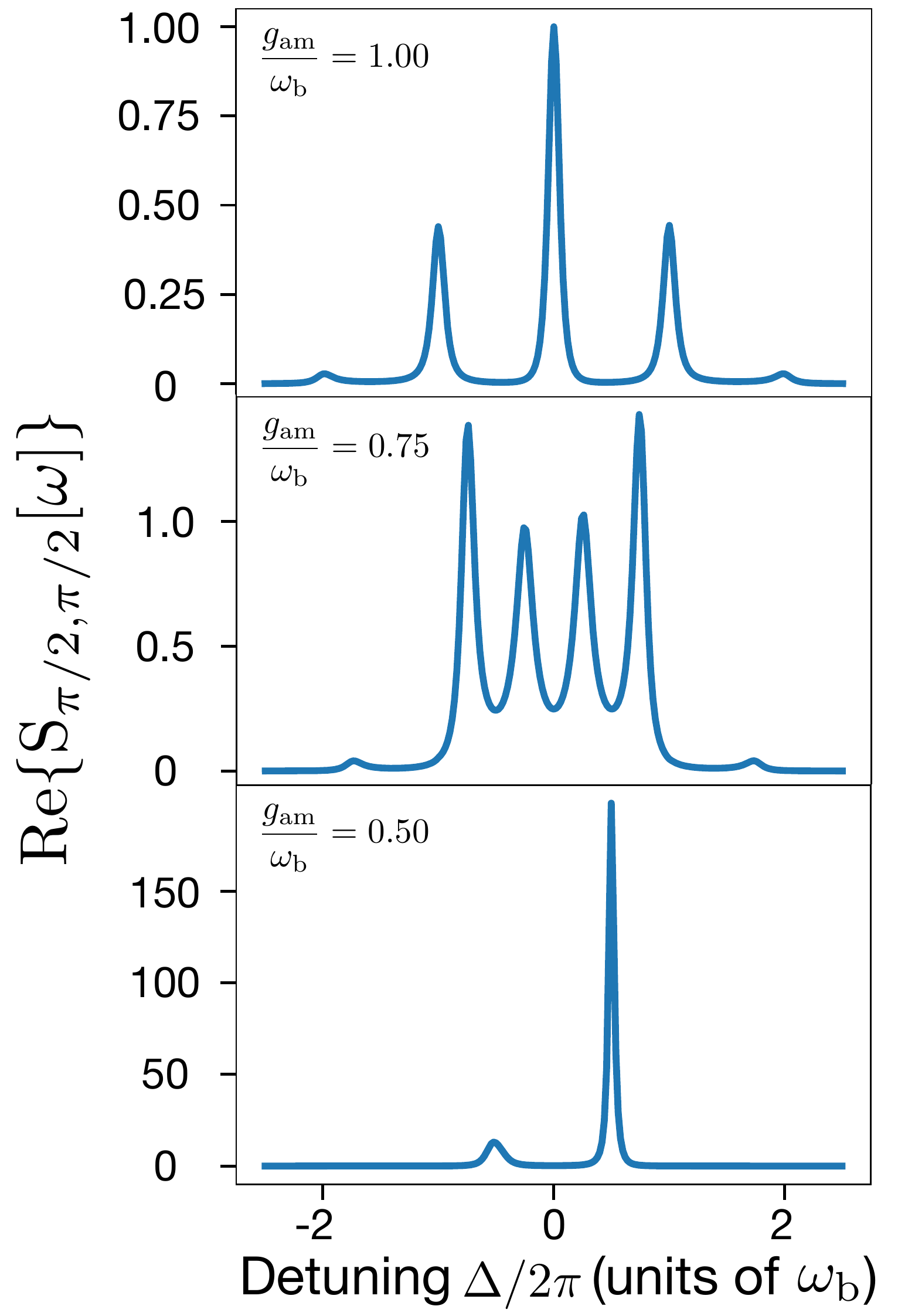}
\caption{Phase-phase autocorrelation spectrum as function of the drive-cavity detuning for $\omega=\omega_{\rm{b}}$ and several values of $g_{\rm{am}}/\omega_{\rm{b}}$.
}\label{Fig:App03}
\end{figure}

In principle the values $g_{\rm{am}}$ and $\omega_{\rm{b}}$ are independent and careful engineering is required to ensure the condition $g_{\rm{am}}=\omega_{\rm{b}}$. The structure of the magnonic spring effect and the magnomechanical damping are strongly dependent on the ratio $g_{\rm{am}}/\omega_{\rm{b}}$. Figs.~\ref{Fig:App01} and \ref{Fig:App02}---the phonon frequency shift and magnomechanical cooling rate, see Eq.~\eqref{eqn:MagSpring}---show this dependence for various values of $g_{\rm{am}}/\omega_{\rm{b}}$. Furthermore, we show in Fig.~\ref{Fig:App03} the phase-phase autocorrelation function for $\omega=\omega_{\rm{b}}$ for these rations of $g_{\rm{am}}/\omega_{\rm{b}}$.

As in the ideal case, the phonon frequency shift and the magnomechanical decay rate vanish at $\Delta =0$. Nevertheless, we see that for $g_{\rm{am}} \neq \omega_{\rm{b}}$ a small deviation from zero detuning can generate a relatively large effect in the phonon frequency and decay rate. This is particularly evident for the case of $g_{\rm{am}}=0.75 \, \omega_{\rm{b}}$. On the other hand, the plots depicting the phase-phase autocorrelation show how the ideal condition $g_{\rm{am}} = \omega_{\rm{b}}$ generates an enhancement in the signal for $\Delta = 0$. For deviations of this condition, the sidebands of the magnon-phonon hybrid modes do not completely interfere. In fact, as depicted in Fig.~\ref{Fig:App03}, the phase-phase autocorrelation at $\Delta = 0$ gets smaller for $g_{\rm{am}} \neq \omega_{\rm{b}}$ and when the sidebands are well separated, as  in the case $g_{\rm{am}} = 0.5 \omega_{\rm{b}}$, the value of the correlation vanishes.

\FloatBarrier

\bibliography{aipsamp}

\begin{thebibliography}{52}%
\makeatletter
\providecommand \@ifxundefined [1]{%
 \@ifx{#1\undefined}
}%
\providecommand \@ifnum [1]{%
 \ifnum #1\expandafter \@firstoftwo
 \else \expandafter \@secondoftwo
 \fi
}%
\providecommand \@ifx [1]{%
 \ifx #1\expandafter \@firstoftwo
 \else \expandafter \@secondoftwo
 \fi
}%
\providecommand \natexlab [1]{#1}%
\providecommand \enquote  [1]{``#1''}%
\providecommand \bibnamefont  [1]{#1}%
\providecommand \bibfnamefont [1]{#1}%
\providecommand \citenamefont [1]{#1}%
\providecommand \href@noop [0]{\@secondoftwo}%
\providecommand \href [0]{\begingroup \@sanitize@url \@href}%
\providecommand \@href[1]{\@@startlink{#1}\@@href}%
\providecommand \@@href[1]{\endgroup#1\@@endlink}%
\providecommand \@sanitize@url [0]{\catcode `\\12\catcode `\$12\catcode
  `\&12\catcode `\#12\catcode `\^12\catcode `\_12\catcode `\%12\relax}%
\providecommand \@@startlink[1]{}%
\providecommand \@@endlink[0]{}%
\providecommand \url  [0]{\begingroup\@sanitize@url \@url }%
\providecommand \@url [1]{\endgroup\@href {#1}{\urlprefix }}%
\providecommand \urlprefix  [0]{URL }%
\providecommand \Eprint [0]{\href }%
\providecommand \doibase [0]{http://dx.doi.org/}%
\providecommand \selectlanguage [0]{\@gobble}%
\providecommand \bibinfo  [0]{\@secondoftwo}%
\providecommand \bibfield  [0]{\@secondoftwo}%
\providecommand \translation [1]{[#1]}%
\providecommand \BibitemOpen [0]{}%
\providecommand \bibitemStop [0]{}%
\providecommand \bibitemNoStop [0]{.\EOS\space}%
\providecommand \EOS [0]{\spacefactor3000\relax}%
\providecommand \BibitemShut  [1]{\csname bibitem#1\endcsname}%
\let\auto@bib@innerbib\@empty
\bibitem [{\citenamefont {{Yeager}}\ and\ \citenamefont
  {{Courts}}(2001)}]{yeager_thermometry_2001}%
  \BibitemOpen
  \bibfield  {author} {\bibinfo {author} {\bibfnamefont {C.~J.}\ \bibnamefont
  {{Yeager}}}\ and\ \bibinfo {author} {\bibfnamefont {S.~S.}\ \bibnamefont
  {{Courts}}},\ }\href@noop {} {\bibfield  {journal} {\bibinfo  {journal} {IEEE
  Sensors Journal}\ }\textbf {\bibinfo {volume} {1}},\ \bibinfo {pages} {352}
  (\bibinfo {year} {2001})}\BibitemShut {NoStop}%
\bibitem [{\citenamefont
  {Preston-Thomas}(1990)}]{Preston-Thomas_International_1989}%
  \BibitemOpen
  \bibfield  {author} {\bibinfo {author} {\bibfnamefont {H.}~\bibnamefont
  {Preston-Thomas}},\ }\href@noop {} {\bibfield  {journal} {\bibinfo  {journal}
  {Metrologia}\ }\textbf {\bibinfo {volume} {27}},\ \bibinfo {pages} {3}
  (\bibinfo {year} {1990})}\BibitemShut {NoStop}%
\bibitem [{\citenamefont {Rusby}\ \emph {et~al.}(2002)\citenamefont {Rusby},
  \citenamefont {Durieux}, \citenamefont {Reesink}, \citenamefont {Hudson},
  \citenamefont {Schuster}, \citenamefont {Kühne}, \citenamefont {Fogle},
  \citenamefont {Soulen},\ and\ \citenamefont
  {Adams}}]{Rusby_Provisional_2002}%
  \BibitemOpen
  \bibfield  {author} {\bibinfo {author} {\bibfnamefont {R.~L.}\ \bibnamefont
  {Rusby}}, \bibinfo {author} {\bibfnamefont {M.}~\bibnamefont {Durieux}},
  \bibinfo {author} {\bibfnamefont {A.~L.}\ \bibnamefont {Reesink}}, \bibinfo
  {author} {\bibfnamefont {R.~P.}\ \bibnamefont {Hudson}}, \bibinfo {author}
  {\bibfnamefont {G.}~\bibnamefont {Schuster}}, \bibinfo {author}
  {\bibfnamefont {M.}~\bibnamefont {Kühne}}, \bibinfo {author} {\bibfnamefont
  {W.~E.}\ \bibnamefont {Fogle}}, \bibinfo {author} {\bibfnamefont {R.~J.}\
  \bibnamefont {Soulen}}, \ and\ \bibinfo {author} {\bibfnamefont {E.~D.}\
  \bibnamefont {Adams}},\ }\href {\doibase 10.1023/A:1013791823354} {\bibfield
  {journal} {\bibinfo  {journal} {Journal of Low Temperature Physics}\ }\textbf
  {\bibinfo {volume} {126}},\ \bibinfo {pages} {633} (\bibinfo {year}
  {2002})}\BibitemShut {NoStop}%
\bibitem [{\citenamefont {Rusby}\ \emph {et~al.}(2007)\citenamefont {Rusby},
  \citenamefont {Fellmuth}, \citenamefont {Engert}, \citenamefont {Fogle},
  \citenamefont {Adams}, \citenamefont {Pitre},\ and\ \citenamefont
  {Durieux}}]{Rusby_Realization_2007}%
  \BibitemOpen
  \bibfield  {author} {\bibinfo {author} {\bibfnamefont {R.~L.}\ \bibnamefont
  {Rusby}}, \bibinfo {author} {\bibfnamefont {B.}~\bibnamefont {Fellmuth}},
  \bibinfo {author} {\bibfnamefont {J.}~\bibnamefont {Engert}}, \bibinfo
  {author} {\bibfnamefont {W.~E.}\ \bibnamefont {Fogle}}, \bibinfo {author}
  {\bibfnamefont {E.~D.}\ \bibnamefont {Adams}}, \bibinfo {author}
  {\bibfnamefont {L.}~\bibnamefont {Pitre}}, \ and\ \bibinfo {author}
  {\bibfnamefont {M.}~\bibnamefont {Durieux}},\ }\href {\doibase
  10.1007/s10909-007-9502-} {\bibfield  {journal} {\bibinfo  {journal} {Journal
  of Low Temperature Physics}\ }\textbf {\bibinfo {volume} {149}},\ \bibinfo
  {pages} {156} (\bibinfo {year} {2007})}\BibitemShut {NoStop}%
\bibitem [{\citenamefont {Greywall}\ and\ \citenamefont
  {Busch}(1982)}]{greywall_3He_1982}%
  \BibitemOpen
  \bibfield  {author} {\bibinfo {author} {\bibfnamefont {D.~S.}\ \bibnamefont
  {Greywall}}\ and\ \bibinfo {author} {\bibfnamefont {P.~A.}\ \bibnamefont
  {Busch}},\ }\href@noop {} {\bibfield  {journal} {\bibinfo  {journal} {Journal
  of Low Temperature Physics}\ }\textbf {\bibinfo {volume} {46}},\ \bibinfo
  {pages} {451} (\bibinfo {year} {1982})}\BibitemShut {NoStop}%
\bibitem [{\citenamefont {Berglund}\ \emph {et~al.}(1972)\citenamefont
  {Berglund}, \citenamefont {Collan}, \citenamefont {Ehnholm}, \citenamefont
  {Gylling},\ and\ \citenamefont {Lounasmaa}}]{berglund_nuclear_1972}%
  \BibitemOpen
  \bibfield  {author} {\bibinfo {author} {\bibfnamefont {P.~M.}\ \bibnamefont
  {Berglund}}, \bibinfo {author} {\bibfnamefont {H.~K.}\ \bibnamefont
  {Collan}}, \bibinfo {author} {\bibfnamefont {G.~J.}\ \bibnamefont {Ehnholm}},
  \bibinfo {author} {\bibfnamefont {R.~G.}\ \bibnamefont {Gylling}}, \ and\
  \bibinfo {author} {\bibfnamefont {O.~V.}\ \bibnamefont {Lounasmaa}},\
  }\href@noop {} {\bibfield  {journal} {\bibinfo  {journal} {Journal of Low
  Temperature Physics}\ }\textbf {\bibinfo {volume} {6}},\ \bibinfo {pages}
  {357} (\bibinfo {year} {1972})}\BibitemShut {NoStop}%
\bibitem [{\citenamefont {Pollanen}\ \emph {et~al.}(2009)\citenamefont
  {Pollanen}, \citenamefont {Choi}, \citenamefont {Davis}, \citenamefont
  {Rolfs},\ and\ \citenamefont {Halperin}}]{Pollanen_Resistance_2009}%
  \BibitemOpen
  \bibfield  {author} {\bibinfo {author} {\bibfnamefont {J.}~\bibnamefont
  {Pollanen}}, \bibinfo {author} {\bibfnamefont {H.}~\bibnamefont {Choi}},
  \bibinfo {author} {\bibfnamefont {J.~P.}\ \bibnamefont {Davis}}, \bibinfo
  {author} {\bibfnamefont {B.~T.}\ \bibnamefont {Rolfs}}, \ and\ \bibinfo
  {author} {\bibfnamefont {W.~P.}\ \bibnamefont {Halperin}},\ }\href {\doibase
  10.1088/1742-6596/150/1/012037} {\bibfield  {journal} {\bibinfo  {journal}
  {Journal of Physics: Conference Series}\ }\textbf {\bibinfo {volume} {150}},\
  \bibinfo {pages} {012037} (\bibinfo {year} {2009})}\BibitemShut {NoStop}%
\bibitem [{\citenamefont {Johnson}(1928)}]{Johnson_Thermal_1928}%
  \BibitemOpen
  \bibfield  {author} {\bibinfo {author} {\bibfnamefont {J.~B.}\ \bibnamefont
  {Johnson}},\ }\href@noop {} {\bibfield  {journal} {\bibinfo  {journal} {Phys.
  Rev.}\ }\textbf {\bibinfo {volume} {32}},\ \bibinfo {pages} {97} (\bibinfo
  {year} {1928})}\BibitemShut {NoStop}%
\bibitem [{\citenamefont {Nyquist}(1928)}]{Nyquist_Thermal_1928}%
  \BibitemOpen
  \bibfield  {author} {\bibinfo {author} {\bibfnamefont {H.}~\bibnamefont
  {Nyquist}},\ }\href {\doibase 10.1103/PhysRev.32.110} {\bibfield  {journal}
  {\bibinfo  {journal} {Phys. Rev.}\ }\textbf {\bibinfo {volume} {32}},\
  \bibinfo {pages} {110} (\bibinfo {year} {1928})}\BibitemShut {NoStop}%
\bibitem [{\citenamefont {Kamper}\ and\ \citenamefont
  {Zimmerman}(1971)}]{Kamper_Noise_1971}%
  \BibitemOpen
  \bibfield  {author} {\bibinfo {author} {\bibfnamefont {R.~A.}\ \bibnamefont
  {Kamper}}\ and\ \bibinfo {author} {\bibfnamefont {J.~E.}\ \bibnamefont
  {Zimmerman}},\ }\href@noop {} {\bibfield  {journal} {\bibinfo  {journal}
  {Journal of Applied Physics}\ }\textbf {\bibinfo {volume} {42}},\ \bibinfo
  {pages} {132} (\bibinfo {year} {1971})}\BibitemShut {NoStop}%
\bibitem [{\citenamefont {Rothfuss}\ \emph {et~al.}(2013)\citenamefont
  {Rothfuss}, \citenamefont {Reiser}, \citenamefont {Fleischmann},\ and\
  \citenamefont {Enss}}]{rothfuss_Noise_2013}%
  \BibitemOpen
  \bibfield  {author} {\bibinfo {author} {\bibfnamefont {D.}~\bibnamefont
  {Rothfuss}}, \bibinfo {author} {\bibfnamefont {A.}~\bibnamefont {Reiser}},
  \bibinfo {author} {\bibfnamefont {A.}~\bibnamefont {Fleischmann}}, \ and\
  \bibinfo {author} {\bibfnamefont {C.}~\bibnamefont {Enss}},\ }\href@noop {}
  {\bibfield  {journal} {\bibinfo  {journal} {Applied Physics Letters}\
  }\textbf {\bibinfo {volume} {103}},\ \bibinfo {pages} {052605} (\bibinfo
  {year} {2013})}\BibitemShut {NoStop}%
\bibitem [{\citenamefont {Rothfuss}\ \emph {et~al.}(2016)\citenamefont
  {Rothfuss}, \citenamefont {Reiser}, \citenamefont {Fleischmann},\ and\
  \citenamefont {Enss}}]{Rothfuss_Noise_2016}%
  \BibitemOpen
  \bibfield  {author} {\bibinfo {author} {\bibfnamefont {D.}~\bibnamefont
  {Rothfuss}}, \bibinfo {author} {\bibfnamefont {A.}~\bibnamefont {Reiser}},
  \bibinfo {author} {\bibfnamefont {A.}~\bibnamefont {Fleischmann}}, \ and\
  \bibinfo {author} {\bibfnamefont {C.}~\bibnamefont {Enss}},\ }\href@noop {}
  {\bibfield  {journal} {\bibinfo  {journal} {Philosophical Transactions of the
  Royal Society A: Mathematical, Physical and Engineering Sciences}\ }\textbf
  {\bibinfo {volume} {374}},\ \bibinfo {pages} {20150051} (\bibinfo {year}
  {2016})}\BibitemShut {NoStop}%
\bibitem [{\citenamefont {Shibahara}\ \emph {et~al.}(2016)\citenamefont
  {Shibahara}, \citenamefont {Hahtela}, \citenamefont {Engert}, \citenamefont
  {van~der Vliet}, \citenamefont {Levitin}, \citenamefont {Casey},
  \citenamefont {Lusher}, \citenamefont {Saunders}, \citenamefont {Drung},\
  and\ \citenamefont {Schurig}}]{Shibahara_Primary_2016}%
  \BibitemOpen
  \bibfield  {author} {\bibinfo {author} {\bibfnamefont {A.}~\bibnamefont
  {Shibahara}}, \bibinfo {author} {\bibfnamefont {O.}~\bibnamefont {Hahtela}},
  \bibinfo {author} {\bibfnamefont {J.}~\bibnamefont {Engert}}, \bibinfo
  {author} {\bibfnamefont {H.}~\bibnamefont {van~der Vliet}}, \bibinfo {author}
  {\bibfnamefont {L.~V.}\ \bibnamefont {Levitin}}, \bibinfo {author}
  {\bibfnamefont {A.}~\bibnamefont {Casey}}, \bibinfo {author} {\bibfnamefont
  {C.~P.}\ \bibnamefont {Lusher}}, \bibinfo {author} {\bibfnamefont
  {J.}~\bibnamefont {Saunders}}, \bibinfo {author} {\bibfnamefont
  {D.}~\bibnamefont {Drung}}, \ and\ \bibinfo {author} {\bibfnamefont
  {T.}~\bibnamefont {Schurig}},\ }\href@noop {} {\bibfield  {journal} {\bibinfo
   {journal} {Philosophical Transactions of the Royal Society A: Mathematical,
  Physical and Engineering Sciences}\ }\textbf {\bibinfo {volume} {374}},\
  \bibinfo {pages} {20150054} (\bibinfo {year} {2016})}\BibitemShut {NoStop}%
\bibitem [{\citenamefont {Hauer}\ \emph {et~al.}(2013)\citenamefont {Hauer},
  \citenamefont {Doolin}, \citenamefont {Beach},\ and\ \citenamefont
  {Davis}}]{Hauer_General_2013}%
  \BibitemOpen
  \bibfield  {author} {\bibinfo {author} {\bibfnamefont {B.}~\bibnamefont
  {Hauer}}, \bibinfo {author} {\bibfnamefont {C.}~\bibnamefont {Doolin}},
  \bibinfo {author} {\bibfnamefont {K.}~\bibnamefont {Beach}}, \ and\ \bibinfo
  {author} {\bibfnamefont {J.}~\bibnamefont {Davis}},\ }\href {\doibase
  https://doi.org/10.1016/j.aop.2013.08.003} {\bibfield  {journal} {\bibinfo
  {journal} {Annals of Physics}\ }\textbf {\bibinfo {volume} {339}},\ \bibinfo
  {pages} {181 } (\bibinfo {year} {2013})}\BibitemShut {NoStop}%
\bibitem [{\citenamefont {Aspelmeyer}\ \emph {et~al.}(2014)\citenamefont
  {Aspelmeyer}, \citenamefont {Kippenberg},\ and\ \citenamefont
  {Marquardt}}]{aspelmeyer_cavity_2014}%
  \BibitemOpen
  \bibfield  {author} {\bibinfo {author} {\bibfnamefont {M.}~\bibnamefont
  {Aspelmeyer}}, \bibinfo {author} {\bibfnamefont {T.~J.}\ \bibnamefont
  {Kippenberg}}, \ and\ \bibinfo {author} {\bibfnamefont {F.}~\bibnamefont
  {Marquardt}},\ }\href {\doibase 10.1103/RevModPhys.86.1391} {\bibfield
  {journal} {\bibinfo  {journal} {Rev. Mod. Phys.}\ }\textbf {\bibinfo {volume}
  {86}},\ \bibinfo {pages} {1391} (\bibinfo {year} {2014})}\BibitemShut
  {NoStop}%
\bibitem [{\citenamefont {MacDonald}\ \emph {et~al.}(2016)\citenamefont
  {MacDonald}, \citenamefont {Hauer}, \citenamefont {Rojas}, \citenamefont
  {Kim}, \citenamefont {Popowich},\ and\ \citenamefont
  {Davis}}]{MacDonald_Optomechanics_2016}%
  \BibitemOpen
  \bibfield  {author} {\bibinfo {author} {\bibfnamefont {A.~J.~R.}\
  \bibnamefont {MacDonald}}, \bibinfo {author} {\bibfnamefont {B.~D.}\
  \bibnamefont {Hauer}}, \bibinfo {author} {\bibfnamefont {X.}~\bibnamefont
  {Rojas}}, \bibinfo {author} {\bibfnamefont {P.~H.}\ \bibnamefont {Kim}},
  \bibinfo {author} {\bibfnamefont {G.~G.}\ \bibnamefont {Popowich}}, \ and\
  \bibinfo {author} {\bibfnamefont {J.~P.}\ \bibnamefont {Davis}},\ }\href
  {\doibase 10.1103/PhysRevA.93.013836} {\bibfield  {journal} {\bibinfo
  {journal} {Phys. Rev. A}\ }\textbf {\bibinfo {volume} {93}},\ \bibinfo
  {pages} {013836} (\bibinfo {year} {2016})}\BibitemShut {NoStop}%
\bibitem [{\citenamefont {Gorodetksy}\ \emph {et~al.}(2010)\citenamefont
  {Gorodetksy}, \citenamefont {Schliesser}, \citenamefont {Anetsberger},
  \citenamefont {Deleglise},\ and\ \citenamefont
  {Kippenberg}}]{Gorodetksy_Phase_2010}%
  \BibitemOpen
  \bibfield  {author} {\bibinfo {author} {\bibfnamefont {M.~L.}\ \bibnamefont
  {Gorodetksy}}, \bibinfo {author} {\bibfnamefont {A.}~\bibnamefont
  {Schliesser}}, \bibinfo {author} {\bibfnamefont {G.}~\bibnamefont
  {Anetsberger}}, \bibinfo {author} {\bibfnamefont {S.}~\bibnamefont
  {Deleglise}}, \ and\ \bibinfo {author} {\bibfnamefont {T.~J.}\ \bibnamefont
  {Kippenberg}},\ }\href {\doibase 10.1364/OE.18.023236} {\bibfield  {journal}
  {\bibinfo  {journal} {Opt. Express}\ }\textbf {\bibinfo {volume} {18}},\
  \bibinfo {pages} {23236} (\bibinfo {year} {2010})}\BibitemShut {NoStop}%
\bibitem [{\citenamefont {B\o{}rkje}\ \emph {et~al.}(2010)\citenamefont
  {B\o{}rkje}, \citenamefont {Nunnenkamp}, \citenamefont {Zwickl},
  \citenamefont {Yang}, \citenamefont {Harris},\ and\ \citenamefont
  {Girvin}}]{borkje_observability_2010}%
  \BibitemOpen
  \bibfield  {author} {\bibinfo {author} {\bibfnamefont {K.}~\bibnamefont
  {B\o{}rkje}}, \bibinfo {author} {\bibfnamefont {A.}~\bibnamefont
  {Nunnenkamp}}, \bibinfo {author} {\bibfnamefont {B.~M.}\ \bibnamefont
  {Zwickl}}, \bibinfo {author} {\bibfnamefont {C.}~\bibnamefont {Yang}},
  \bibinfo {author} {\bibfnamefont {J.~G.~E.}\ \bibnamefont {Harris}}, \ and\
  \bibinfo {author} {\bibfnamefont {S.~M.}\ \bibnamefont {Girvin}},\ }\href
  {\doibase 10.1103/PhysRevA.82.013818} {\bibfield  {journal} {\bibinfo
  {journal} {Phys. Rev. A}\ }\textbf {\bibinfo {volume} {82}},\ \bibinfo
  {pages} {013818} (\bibinfo {year} {2010})}\BibitemShut {NoStop}%
\bibitem [{\citenamefont {Purdy}\ \emph {et~al.}(2017)\citenamefont {Purdy},
  \citenamefont {Grutter}, \citenamefont {Srinivasan},\ and\ \citenamefont
  {Taylor}}]{purdy_quantum_2017}%
  \BibitemOpen
  \bibfield  {author} {\bibinfo {author} {\bibfnamefont {T.~P.}\ \bibnamefont
  {Purdy}}, \bibinfo {author} {\bibfnamefont {K.~E.}\ \bibnamefont {Grutter}},
  \bibinfo {author} {\bibfnamefont {K.}~\bibnamefont {Srinivasan}}, \ and\
  \bibinfo {author} {\bibfnamefont {J.~M.}\ \bibnamefont {Taylor}},\ }\href
  {\doibase 10.1126/science.aag1407} {\bibfield  {journal} {\bibinfo  {journal}
  {Science}\ }\textbf {\bibinfo {volume} {356}},\ \bibinfo {pages} {1265}
  (\bibinfo {year} {2017})}\BibitemShut {NoStop}%
\bibitem [{\citenamefont {Meenehan}\ \emph {et~al.}(2015)\citenamefont
  {Meenehan}, \citenamefont {Cohen}, \citenamefont {MacCabe}, \citenamefont
  {Marsili}, \citenamefont {Shaw},\ and\ \citenamefont
  {Painter}}]{Meenehan_Pulsed_2015}%
  \BibitemOpen
  \bibfield  {author} {\bibinfo {author} {\bibfnamefont {S.~M.}\ \bibnamefont
  {Meenehan}}, \bibinfo {author} {\bibfnamefont {J.~D.}\ \bibnamefont {Cohen}},
  \bibinfo {author} {\bibfnamefont {G.~S.}\ \bibnamefont {MacCabe}}, \bibinfo
  {author} {\bibfnamefont {F.}~\bibnamefont {Marsili}}, \bibinfo {author}
  {\bibfnamefont {M.~D.}\ \bibnamefont {Shaw}}, \ and\ \bibinfo {author}
  {\bibfnamefont {O.}~\bibnamefont {Painter}},\ }\href {\doibase
  10.1103/PhysRevX.5.041002} {\bibfield  {journal} {\bibinfo  {journal} {Phys.
  Rev. X}\ }\textbf {\bibinfo {volume} {5}},\ \bibinfo {pages} {041002}
  (\bibinfo {year} {2015})}\BibitemShut {NoStop}%
\bibitem [{\citenamefont {Hauer}\ \emph {et~al.}(2018)\citenamefont {Hauer},
  \citenamefont {Kim}, \citenamefont {Doolin}, \citenamefont {Souris},\ and\
  \citenamefont {Davis}}]{hauer_two-level_2018}%
  \BibitemOpen
  \bibfield  {author} {\bibinfo {author} {\bibfnamefont {B.~D.}\ \bibnamefont
  {Hauer}}, \bibinfo {author} {\bibfnamefont {P.~H.}\ \bibnamefont {Kim}},
  \bibinfo {author} {\bibfnamefont {C.}~\bibnamefont {Doolin}}, \bibinfo
  {author} {\bibfnamefont {F.}~\bibnamefont {Souris}}, \ and\ \bibinfo {author}
  {\bibfnamefont {J.~P.}\ \bibnamefont {Davis}},\ }\href {\doibase
  10.1103/PhysRevB.98.214303} {\bibfield  {journal} {\bibinfo  {journal} {Phys.
  Rev. B}\ }\textbf {\bibinfo {volume} {98}},\ \bibinfo {pages} {214303}
  (\bibinfo {year} {2018})}\BibitemShut {NoStop}%
\bibitem [{\citenamefont {Ramp}\ \emph {et~al.}(2019)\citenamefont {Ramp},
  \citenamefont {Hauer}, \citenamefont {Balram}, \citenamefont {Clark},
  \citenamefont {Srinivasan},\ and\ \citenamefont
  {Davis}}]{ramp_elimination_2019}%
  \BibitemOpen
  \bibfield  {author} {\bibinfo {author} {\bibfnamefont {H.}~\bibnamefont
  {Ramp}}, \bibinfo {author} {\bibfnamefont {B.~D.}\ \bibnamefont {Hauer}},
  \bibinfo {author} {\bibfnamefont {K.~C.}\ \bibnamefont {Balram}}, \bibinfo
  {author} {\bibfnamefont {T.~J.}\ \bibnamefont {Clark}}, \bibinfo {author}
  {\bibfnamefont {K.}~\bibnamefont {Srinivasan}}, \ and\ \bibinfo {author}
  {\bibfnamefont {J.~P.}\ \bibnamefont {Davis}},\ }\href@noop {} {\bibfield
  {journal} {\bibinfo  {journal} {Phys. Rev. Lett.}\ }\textbf {\bibinfo
  {volume} {123}},\ \bibinfo {pages} {093603} (\bibinfo {year}
  {2019})}\BibitemShut {NoStop}%
\bibitem [{\citenamefont {Woolley}\ \emph {et~al.}(2008)\citenamefont
  {Woolley}, \citenamefont {Doherty}, \citenamefont {Milburn},\ and\
  \citenamefont {Schwab}}]{Woolley_Nanomechanical_2008}%
  \BibitemOpen
  \bibfield  {author} {\bibinfo {author} {\bibfnamefont {M.~J.}\ \bibnamefont
  {Woolley}}, \bibinfo {author} {\bibfnamefont {A.~C.}\ \bibnamefont
  {Doherty}}, \bibinfo {author} {\bibfnamefont {G.~J.}\ \bibnamefont
  {Milburn}}, \ and\ \bibinfo {author} {\bibfnamefont {K.~C.}\ \bibnamefont
  {Schwab}},\ }\href {\doibase 10.1103/PhysRevA.78.062303} {\bibfield
  {journal} {\bibinfo  {journal} {Phys. Rev. A}\ }\textbf {\bibinfo {volume}
  {78}},\ \bibinfo {pages} {062303} (\bibinfo {year} {2008})}\BibitemShut
  {NoStop}%
\bibitem [{\citenamefont {Regal}\ \emph {et~al.}(2008)\citenamefont {Regal},
  \citenamefont {Teufel},\ and\ \citenamefont
  {Lehnert}}]{Regal_Measuring_2008}%
  \BibitemOpen
  \bibfield  {author} {\bibinfo {author} {\bibfnamefont {C.}~\bibnamefont
  {Regal}}, \bibinfo {author} {\bibfnamefont {J.}~\bibnamefont {Teufel}}, \
  and\ \bibinfo {author} {\bibfnamefont {K.}~\bibnamefont {Lehnert}},\ }\href
  {\doibase 10.1038/nphys974} {\bibfield  {journal} {\bibinfo  {journal}
  {Nature Physics}\ }\textbf {\bibinfo {volume} {4}},\ \bibinfo {pages} {555}
  (\bibinfo {year} {2008})}\BibitemShut {NoStop}%
\bibitem [{\citenamefont {Palomaki}\ \emph {et~al.}(2013)\citenamefont
  {Palomaki}, \citenamefont {Teufel}, \citenamefont {Simmonds},\ and\
  \citenamefont {Lehnert}}]{Palomaki_Entangling_2013}%
  \BibitemOpen
  \bibfield  {author} {\bibinfo {author} {\bibfnamefont {T.~A.}\ \bibnamefont
  {Palomaki}}, \bibinfo {author} {\bibfnamefont {J.~D.}\ \bibnamefont
  {Teufel}}, \bibinfo {author} {\bibfnamefont {R.~W.}\ \bibnamefont
  {Simmonds}}, \ and\ \bibinfo {author} {\bibfnamefont {K.~W.}\ \bibnamefont
  {Lehnert}},\ }\href {\doibase 10.1126/science.1244563} {\bibfield  {journal}
  {\bibinfo  {journal} {Science}\ }\textbf {\bibinfo {volume} {342}},\ \bibinfo
  {pages} {710} (\bibinfo {year} {2013})}\BibitemShut {NoStop}%
\bibitem [{\citenamefont {Suh}\ \emph {et~al.}(2014)\citenamefont {Suh},
  \citenamefont {Weinstein}, \citenamefont {Lei}, \citenamefont {Wollman},
  \citenamefont {Steinke}, \citenamefont {Meystre}, \citenamefont {Clerk},\
  and\ \citenamefont {Schwab}}]{Suh_Mechanically_2014}%
  \BibitemOpen
  \bibfield  {author} {\bibinfo {author} {\bibfnamefont {J.}~\bibnamefont
  {Suh}}, \bibinfo {author} {\bibfnamefont {A.~J.}\ \bibnamefont {Weinstein}},
  \bibinfo {author} {\bibfnamefont {C.~U.}\ \bibnamefont {Lei}}, \bibinfo
  {author} {\bibfnamefont {E.~E.}\ \bibnamefont {Wollman}}, \bibinfo {author}
  {\bibfnamefont {S.~K.}\ \bibnamefont {Steinke}}, \bibinfo {author}
  {\bibfnamefont {P.}~\bibnamefont {Meystre}}, \bibinfo {author} {\bibfnamefont
  {A.~A.}\ \bibnamefont {Clerk}}, \ and\ \bibinfo {author} {\bibfnamefont
  {K.~C.}\ \bibnamefont {Schwab}},\ }\href {\doibase 10.1126/science.1253258}
  {\bibfield  {journal} {\bibinfo  {journal} {Science}\ }\textbf {\bibinfo
  {volume} {344}},\ \bibinfo {pages} {1262} (\bibinfo {year}
  {2014})}\BibitemShut {NoStop}%
\bibitem [{\citenamefont {Pirkkalainen}\ \emph {et~al.}(2015)\citenamefont
  {Pirkkalainen}, \citenamefont {Damsk\"agg}, \citenamefont {Brandt},
  \citenamefont {Massel},\ and\ \citenamefont
  {Sillanp\"a\"a}}]{Pirkkalainen2015}%
  \BibitemOpen
  \bibfield  {author} {\bibinfo {author} {\bibfnamefont {J.-M.}\ \bibnamefont
  {Pirkkalainen}}, \bibinfo {author} {\bibfnamefont {E.}~\bibnamefont
  {Damsk\"agg}}, \bibinfo {author} {\bibfnamefont {M.}~\bibnamefont {Brandt}},
  \bibinfo {author} {\bibfnamefont {F.}~\bibnamefont {Massel}}, \ and\ \bibinfo
  {author} {\bibfnamefont {M.~A.}\ \bibnamefont {Sillanp\"a\"a}},\ }\href
  {\doibase 10.1103/PhysRevLett.115.243601} {\bibfield  {journal} {\bibinfo
  {journal} {Phys. Rev. Lett.}\ }\textbf {\bibinfo {volume} {115}},\ \bibinfo
  {pages} {243601} (\bibinfo {year} {2015})}\BibitemShut {NoStop}%
\bibitem [{\citenamefont {Lecocq}\ \emph {et~al.}(2016)\citenamefont {Lecocq},
  \citenamefont {Clark}, \citenamefont {Simmonds}, \citenamefont {Aumentado},\
  and\ \citenamefont {Teufel}}]{Lecocq_Mechanically_2016}%
  \BibitemOpen
  \bibfield  {author} {\bibinfo {author} {\bibfnamefont {F.}~\bibnamefont
  {Lecocq}}, \bibinfo {author} {\bibfnamefont {J.~B.}\ \bibnamefont {Clark}},
  \bibinfo {author} {\bibfnamefont {R.~W.}\ \bibnamefont {Simmonds}}, \bibinfo
  {author} {\bibfnamefont {J.}~\bibnamefont {Aumentado}}, \ and\ \bibinfo
  {author} {\bibfnamefont {J.~D.}\ \bibnamefont {Teufel}},\ }\href {\doibase
  10.1103/PhysRevLett.116.043601} {\bibfield  {journal} {\bibinfo  {journal}
  {Phys. Rev. Lett.}\ }\textbf {\bibinfo {volume} {116}},\ \bibinfo {pages}
  {043601} (\bibinfo {year} {2016})}\BibitemShut {NoStop}%
\bibitem [{\citenamefont {Zhang}\ \emph {et~al.}(2014)\citenamefont {Zhang},
  \citenamefont {Zou}, \citenamefont {Jiang},\ and\ \citenamefont
  {Tang}}]{zhang_strongly_2014}%
  \BibitemOpen
  \bibfield  {author} {\bibinfo {author} {\bibfnamefont {X.}~\bibnamefont
  {Zhang}}, \bibinfo {author} {\bibfnamefont {C.-L.}\ \bibnamefont {Zou}},
  \bibinfo {author} {\bibfnamefont {L.}~\bibnamefont {Jiang}}, \ and\ \bibinfo
  {author} {\bibfnamefont {H.~X.}\ \bibnamefont {Tang}},\ }\href {\doibase
  10.1103/PhysRevLett.113.156401} {\bibfield  {journal} {\bibinfo  {journal}
  {Phys. Rev. Lett.}\ }\textbf {\bibinfo {volume} {113}},\ \bibinfo {pages}
  {156401} (\bibinfo {year} {2014})}\BibitemShut {NoStop}%
\bibitem [{\citenamefont {Goryachev}\ \emph {et~al.}(2014)\citenamefont
  {Goryachev}, \citenamefont {Farr}, \citenamefont {Creedon}, \citenamefont
  {Fan}, \citenamefont {Kostylev},\ and\ \citenamefont
  {Tobar}}]{goryachev_high_coop_2014}%
  \BibitemOpen
  \bibfield  {author} {\bibinfo {author} {\bibfnamefont {M.}~\bibnamefont
  {Goryachev}}, \bibinfo {author} {\bibfnamefont {W.~G.}\ \bibnamefont {Farr}},
  \bibinfo {author} {\bibfnamefont {D.~L.}\ \bibnamefont {Creedon}}, \bibinfo
  {author} {\bibfnamefont {Y.}~\bibnamefont {Fan}}, \bibinfo {author}
  {\bibfnamefont {M.}~\bibnamefont {Kostylev}}, \ and\ \bibinfo {author}
  {\bibfnamefont {M.~E.}\ \bibnamefont {Tobar}},\ }\href {\doibase
  10.1103/PhysRevApplied.2.054002} {\bibfield  {journal} {\bibinfo  {journal}
  {Phys. Rev. Applied}\ }\textbf {\bibinfo {volume} {2}},\ \bibinfo {pages}
  {054002} (\bibinfo {year} {2014})}\BibitemShut {NoStop}%
\bibitem [{\citenamefont {Tabuchi}\ \emph {et~al.}(2014)\citenamefont
  {Tabuchi}, \citenamefont {Ishino}, \citenamefont {Ishikawa}, \citenamefont
  {Yamazaki}, \citenamefont {Usami},\ and\ \citenamefont
  {Nakamura}}]{tabuchi_hybridizing_2014}%
  \BibitemOpen
  \bibfield  {author} {\bibinfo {author} {\bibfnamefont {Y.}~\bibnamefont
  {Tabuchi}}, \bibinfo {author} {\bibfnamefont {S.}~\bibnamefont {Ishino}},
  \bibinfo {author} {\bibfnamefont {T.}~\bibnamefont {Ishikawa}}, \bibinfo
  {author} {\bibfnamefont {R.}~\bibnamefont {Yamazaki}}, \bibinfo {author}
  {\bibfnamefont {K.}~\bibnamefont {Usami}}, \ and\ \bibinfo {author}
  {\bibfnamefont {Y.}~\bibnamefont {Nakamura}},\ }\href {\doibase
  10.1103/PhysRevLett.113.083603} {\bibfield  {journal} {\bibinfo  {journal}
  {Phys. Rev. Lett.}\ }\textbf {\bibinfo {volume} {113}},\ \bibinfo {pages}
  {083603} (\bibinfo {year} {2014})}\BibitemShut {NoStop}%
\bibitem [{\citenamefont {Zhang}\ \emph {et~al.}(2016)\citenamefont {Zhang},
  \citenamefont {Zou}, \citenamefont {Jiang},\ and\ \citenamefont
  {Tang}}]{zhang_cavity_2016}%
  \BibitemOpen
  \bibfield  {author} {\bibinfo {author} {\bibfnamefont {X.}~\bibnamefont
  {Zhang}}, \bibinfo {author} {\bibfnamefont {C.-L.}\ \bibnamefont {Zou}},
  \bibinfo {author} {\bibfnamefont {L.}~\bibnamefont {Jiang}}, \ and\ \bibinfo
  {author} {\bibfnamefont {H.~X.}\ \bibnamefont {Tang}},\ }\href {\doibase
  10.1126/sciadv.1501286} {\bibfield  {journal} {\bibinfo  {journal} {Science
  Advances}\ }\textbf {\bibinfo {volume} {2}},\ \bibinfo {pages} {e1501286}
  (\bibinfo {year} {2016})}\BibitemShut {NoStop}%
\bibitem [{\citenamefont {Tabuchi}\ \emph {et~al.}(2015)\citenamefont
  {Tabuchi}, \citenamefont {Ishino}, \citenamefont {Noguchi}, \citenamefont
  {Ishikawa}, \citenamefont {Yamazaki}, \citenamefont {Usami},\ and\
  \citenamefont {Nakamura}}]{tabuchi_coherent_coupling_2015}%
  \BibitemOpen
  \bibfield  {author} {\bibinfo {author} {\bibfnamefont {Y.}~\bibnamefont
  {Tabuchi}}, \bibinfo {author} {\bibfnamefont {S.}~\bibnamefont {Ishino}},
  \bibinfo {author} {\bibfnamefont {A.}~\bibnamefont {Noguchi}}, \bibinfo
  {author} {\bibfnamefont {T.}~\bibnamefont {Ishikawa}}, \bibinfo {author}
  {\bibfnamefont {R.}~\bibnamefont {Yamazaki}}, \bibinfo {author}
  {\bibfnamefont {K.}~\bibnamefont {Usami}}, \ and\ \bibinfo {author}
  {\bibfnamefont {Y.}~\bibnamefont {Nakamura}},\ }\href@noop {} {\bibfield
  {journal} {\bibinfo  {journal} {Science}\ }\textbf {\bibinfo {volume}
  {349}},\ \bibinfo {pages} {405} (\bibinfo {year} {2015})}\BibitemShut
  {NoStop}%
\bibitem [{\citenamefont {Bai}\ \emph {et~al.}(2015)\citenamefont {Bai},
  \citenamefont {Harder}, \citenamefont {Chen}, \citenamefont {Fan},
  \citenamefont {Xiao},\ and\ \citenamefont {Hu}}]{bai_spin_pumping_2015}%
  \BibitemOpen
  \bibfield  {author} {\bibinfo {author} {\bibfnamefont {L.}~\bibnamefont
  {Bai}}, \bibinfo {author} {\bibfnamefont {M.}~\bibnamefont {Harder}},
  \bibinfo {author} {\bibfnamefont {Y.~P.}\ \bibnamefont {Chen}}, \bibinfo
  {author} {\bibfnamefont {X.}~\bibnamefont {Fan}}, \bibinfo {author}
  {\bibfnamefont {J.~Q.}\ \bibnamefont {Xiao}}, \ and\ \bibinfo {author}
  {\bibfnamefont {C.-M.}\ \bibnamefont {Hu}},\ }\href {\doibase
  10.1103/PhysRevLett.114.227201} {\bibfield  {journal} {\bibinfo  {journal}
  {Phys. Rev. Lett.}\ }\textbf {\bibinfo {volume} {114}},\ \bibinfo {pages}
  {227201} (\bibinfo {year} {2015})}\BibitemShut {NoStop}%
\bibitem [{\citenamefont {Viennot}\ \emph {et~al.}(2015)\citenamefont
  {Viennot}, \citenamefont {Dartiailh}, \citenamefont {Cottet},\ and\
  \citenamefont {Kontos}}]{viennot_coherent_coupling_2015}%
  \BibitemOpen
  \bibfield  {author} {\bibinfo {author} {\bibfnamefont {J.~J.}\ \bibnamefont
  {Viennot}}, \bibinfo {author} {\bibfnamefont {M.~C.}\ \bibnamefont
  {Dartiailh}}, \bibinfo {author} {\bibfnamefont {A.}~\bibnamefont {Cottet}}, \
  and\ \bibinfo {author} {\bibfnamefont {T.}~\bibnamefont {Kontos}},\
  }\href@noop {} {\bibfield  {journal} {\bibinfo  {journal} {Science}\ }\textbf
  {\bibinfo {volume} {349}},\ \bibinfo {pages} {408} (\bibinfo {year}
  {2015})}\BibitemShut {NoStop}%
\bibitem [{\citenamefont {Zhang}\ \emph
  {et~al.}(2015{\natexlab{a}})\citenamefont {Zhang}, \citenamefont {Wang},
  \citenamefont {Li}, \citenamefont {Luo}, \citenamefont {Wu}, \citenamefont
  {Nori},\ and\ \citenamefont {You}}]{zhang_cavity_quatum_2015}%
  \BibitemOpen
  \bibfield  {author} {\bibinfo {author} {\bibfnamefont {D.}~\bibnamefont
  {Zhang}}, \bibinfo {author} {\bibfnamefont {X.-M.}\ \bibnamefont {Wang}},
  \bibinfo {author} {\bibfnamefont {T.-F.}\ \bibnamefont {Li}}, \bibinfo
  {author} {\bibfnamefont {X.-Q.}\ \bibnamefont {Luo}}, \bibinfo {author}
  {\bibfnamefont {W.}~\bibnamefont {Wu}}, \bibinfo {author} {\bibfnamefont
  {F.}~\bibnamefont {Nori}}, \ and\ \bibinfo {author} {\bibfnamefont {J.~Q.}\
  \bibnamefont {You}},\ }\href@noop {} {\bibfield  {journal} {\bibinfo
  {journal} {npj Quantum Information}\ }\textbf {\bibinfo {volume} {1}},\
  \bibinfo {pages} {15014} (\bibinfo {year} {2015}{\natexlab{a}})}\BibitemShut
  {NoStop}%
\bibitem [{\citenamefont {Cao}\ \emph {et~al.}(2015)\citenamefont {Cao},
  \citenamefont {Yan}, \citenamefont {Huebl}, \citenamefont {Goennenwein},\
  and\ \citenamefont {Bauer}}]{cao_exchange_2015}%
  \BibitemOpen
  \bibfield  {author} {\bibinfo {author} {\bibfnamefont {Y.}~\bibnamefont
  {Cao}}, \bibinfo {author} {\bibfnamefont {P.}~\bibnamefont {Yan}}, \bibinfo
  {author} {\bibfnamefont {H.}~\bibnamefont {Huebl}}, \bibinfo {author}
  {\bibfnamefont {S.~T.~B.}\ \bibnamefont {Goennenwein}}, \ and\ \bibinfo
  {author} {\bibfnamefont {G.~E.~W.}\ \bibnamefont {Bauer}},\ }\href@noop {}
  {\bibfield  {journal} {\bibinfo  {journal} {Phys. Rev. B}\ }\textbf {\bibinfo
  {volume} {91}},\ \bibinfo {pages} {094423} (\bibinfo {year}
  {2015})}\BibitemShut {NoStop}%
\bibitem [{\citenamefont {Zhang}\ \emph
  {et~al.}(2015{\natexlab{b}})\citenamefont {Zhang}, \citenamefont {Zou},
  \citenamefont {Zhu}, \citenamefont {Marquardt}, \citenamefont {Jiang},\ and\
  \citenamefont {Tang}}]{zhang_magnon_dark_2015}%
  \BibitemOpen
  \bibfield  {author} {\bibinfo {author} {\bibfnamefont {X.}~\bibnamefont
  {Zhang}}, \bibinfo {author} {\bibfnamefont {C.-L.}\ \bibnamefont {Zou}},
  \bibinfo {author} {\bibfnamefont {N.}~\bibnamefont {Zhu}}, \bibinfo {author}
  {\bibfnamefont {F.}~\bibnamefont {Marquardt}}, \bibinfo {author}
  {\bibfnamefont {L.}~\bibnamefont {Jiang}}, \ and\ \bibinfo {author}
  {\bibfnamefont {H.~X.}\ \bibnamefont {Tang}},\ }\href@noop {} {\bibfield
  {journal} {\bibinfo  {journal} {Nature Communications}\ }\textbf {\bibinfo
  {volume} {6}},\ \bibinfo {pages} {8914} (\bibinfo {year}
  {2015}{\natexlab{b}})}\BibitemShut {NoStop}%
\bibitem [{\citenamefont {Hisatomi}\ \emph {et~al.}(2016)\citenamefont
  {Hisatomi}, \citenamefont {Osada}, \citenamefont {Tabuchi}, \citenamefont
  {Ishikawa}, \citenamefont {Noguchi}, \citenamefont {Yamazaki}, \citenamefont
  {Usami},\ and\ \citenamefont
  {Nakamura}}]{hisatomi_bidirectional_conversion_2016}%
  \BibitemOpen
  \bibfield  {author} {\bibinfo {author} {\bibfnamefont {R.}~\bibnamefont
  {Hisatomi}}, \bibinfo {author} {\bibfnamefont {A.}~\bibnamefont {Osada}},
  \bibinfo {author} {\bibfnamefont {Y.}~\bibnamefont {Tabuchi}}, \bibinfo
  {author} {\bibfnamefont {T.}~\bibnamefont {Ishikawa}}, \bibinfo {author}
  {\bibfnamefont {A.}~\bibnamefont {Noguchi}}, \bibinfo {author} {\bibfnamefont
  {R.}~\bibnamefont {Yamazaki}}, \bibinfo {author} {\bibfnamefont
  {K.}~\bibnamefont {Usami}}, \ and\ \bibinfo {author} {\bibfnamefont
  {Y.}~\bibnamefont {Nakamura}},\ }\href@noop {} {\bibfield  {journal}
  {\bibinfo  {journal} {Phys. Rev. B}\ }\textbf {\bibinfo {volume} {93}},\
  \bibinfo {pages} {174427} (\bibinfo {year} {2016})}\BibitemShut {NoStop}%
\bibitem [{\citenamefont {Lachance-Quirion}\ \emph {et~al.}(2017)\citenamefont
  {Lachance-Quirion}, \citenamefont {Tabuchi}, \citenamefont {Ishino},
  \citenamefont {Noguchi}, \citenamefont {Ishikawa}, \citenamefont {Yamazaki},\
  and\ \citenamefont {Nakamura}}]{lachance_resolving_quanta_2017}%
  \BibitemOpen
  \bibfield  {author} {\bibinfo {author} {\bibfnamefont {D.}~\bibnamefont
  {Lachance-Quirion}}, \bibinfo {author} {\bibfnamefont {Y.}~\bibnamefont
  {Tabuchi}}, \bibinfo {author} {\bibfnamefont {S.}~\bibnamefont {Ishino}},
  \bibinfo {author} {\bibfnamefont {A.}~\bibnamefont {Noguchi}}, \bibinfo
  {author} {\bibfnamefont {T.}~\bibnamefont {Ishikawa}}, \bibinfo {author}
  {\bibfnamefont {R.}~\bibnamefont {Yamazaki}}, \ and\ \bibinfo {author}
  {\bibfnamefont {Y.}~\bibnamefont {Nakamura}},\ }\href@noop {} {\bibfield
  {journal} {\bibinfo  {journal} {Science Advances}\ }\textbf {\bibinfo
  {volume} {3}},\ \bibinfo {pages} {e1603150} (\bibinfo {year}
  {2017})}\BibitemShut {NoStop}%
\bibitem [{\citenamefont {Wang}\ \emph {et~al.}(2018)\citenamefont {Wang},
  \citenamefont {Zhang}, \citenamefont {Zhang}, \citenamefont {Li},
  \citenamefont {Hu},\ and\ \citenamefont {You}}]{wang_bistability_2018}%
  \BibitemOpen
  \bibfield  {author} {\bibinfo {author} {\bibfnamefont {Y.-P.}\ \bibnamefont
  {Wang}}, \bibinfo {author} {\bibfnamefont {G.-Q.}\ \bibnamefont {Zhang}},
  \bibinfo {author} {\bibfnamefont {D.}~\bibnamefont {Zhang}}, \bibinfo
  {author} {\bibfnamefont {T.-F.}\ \bibnamefont {Li}}, \bibinfo {author}
  {\bibfnamefont {C.-M.}\ \bibnamefont {Hu}}, \ and\ \bibinfo {author}
  {\bibfnamefont {J.~Q.}\ \bibnamefont {You}},\ }\href@noop {} {\bibfield
  {journal} {\bibinfo  {journal} {Phys. Rev. Lett.}\ }\textbf {\bibinfo
  {volume} {120}},\ \bibinfo {pages} {057202} (\bibinfo {year}
  {2018})}\BibitemShut {NoStop}%
\bibitem [{\citenamefont {Hou}\ and\ \citenamefont
  {Liu}(2019)}]{hou_strong_coupling_2019}%
  \BibitemOpen
  \bibfield  {author} {\bibinfo {author} {\bibfnamefont {J.~T.}\ \bibnamefont
  {Hou}}\ and\ \bibinfo {author} {\bibfnamefont {L.}~\bibnamefont {Liu}},\
  }\href@noop {} {\bibfield  {journal} {\bibinfo  {journal} {Phys. Rev. Lett.}\
  }\textbf {\bibinfo {volume} {123}},\ \bibinfo {pages} {107702} (\bibinfo
  {year} {2019})}\BibitemShut {NoStop}%
\bibitem [{\citenamefont {Lachance-Quirion}\ \emph {et~al.}(2019)\citenamefont
  {Lachance-Quirion}, \citenamefont {Tabuchi}, \citenamefont {Gloppe},
  \citenamefont {Usami},\ and\ \citenamefont
  {Nakamura}}]{Lachance_Hybrid_2019}%
  \BibitemOpen
  \bibfield  {author} {\bibinfo {author} {\bibfnamefont {D.}~\bibnamefont
  {Lachance-Quirion}}, \bibinfo {author} {\bibfnamefont {Y.}~\bibnamefont
  {Tabuchi}}, \bibinfo {author} {\bibfnamefont {A.}~\bibnamefont {Gloppe}},
  \bibinfo {author} {\bibfnamefont {K.}~\bibnamefont {Usami}}, \ and\ \bibinfo
  {author} {\bibfnamefont {Y.}~\bibnamefont {Nakamura}},\ }\href {\doibase
  10.7567/1882-0786/ab248d} {\bibfield  {journal} {\bibinfo  {journal} {Applied
  Physics Express}\ }\textbf {\bibinfo {volume} {12}},\ \bibinfo {pages}
  {070101} (\bibinfo {year} {2019})}\BibitemShut {NoStop}%
\bibitem [{\citenamefont {Wang}\ \emph {et~al.}(2019)\citenamefont {Wang},
  \citenamefont {Rao}, \citenamefont {Yang}, \citenamefont {Xu}, \citenamefont
  {Gui}, \citenamefont {Yao}, \citenamefont {You},\ and\ \citenamefont
  {Hu}}]{Wang_Nonreciprocity_2019}%
  \BibitemOpen
  \bibfield  {author} {\bibinfo {author} {\bibfnamefont {Y.-P.}\ \bibnamefont
  {Wang}}, \bibinfo {author} {\bibfnamefont {J.~W.}\ \bibnamefont {Rao}},
  \bibinfo {author} {\bibfnamefont {Y.}~\bibnamefont {Yang}}, \bibinfo {author}
  {\bibfnamefont {P.-C.}\ \bibnamefont {Xu}}, \bibinfo {author} {\bibfnamefont
  {Y.~S.}\ \bibnamefont {Gui}}, \bibinfo {author} {\bibfnamefont {B.~M.}\
  \bibnamefont {Yao}}, \bibinfo {author} {\bibfnamefont {J.~Q.}\ \bibnamefont
  {You}}, \ and\ \bibinfo {author} {\bibfnamefont {C.-M.}\ \bibnamefont {Hu}},\
  }\href {\doibase 10.1103/PhysRevLett.123.127202} {\bibfield  {journal}
  {\bibinfo  {journal} {Phys. Rev. Lett.}\ }\textbf {\bibinfo {volume} {123}},\
  \bibinfo {pages} {127202} (\bibinfo {year} {2019})}\BibitemShut {NoStop}%
\bibitem [{\citenamefont {Callen}\ and\ \citenamefont
  {Welton}(1951)}]{callen_irreversibility_1951}%
  \BibitemOpen
  \bibfield  {author} {\bibinfo {author} {\bibfnamefont {H.~B.}\ \bibnamefont
  {Callen}}\ and\ \bibinfo {author} {\bibfnamefont {T.~A.}\ \bibnamefont
  {Welton}},\ }\href@noop {} {\bibfield  {journal} {\bibinfo  {journal} {Phys.
  Rev.}\ }\textbf {\bibinfo {volume} {83}},\ \bibinfo {pages} {34} (\bibinfo
  {year} {1951})}\BibitemShut {NoStop}%
\bibitem [{\citenamefont {Kubo}(1966)}]{kubo_the_fluctuation_1966}%
  \BibitemOpen
  \bibfield  {author} {\bibinfo {author} {\bibfnamefont {R.}~\bibnamefont
  {Kubo}},\ }\href {\doibase 10.1088/0034-4885/29/1/306} {\bibfield  {journal}
  {\bibinfo  {journal} {Reports on Progress in Physics}\ }\textbf {\bibinfo
  {volume} {29}},\ \bibinfo {pages} {255} (\bibinfo {year} {1966})}\BibitemShut
  {NoStop}%
\bibitem [{\citenamefont {Gardiner}\ and\ \citenamefont
  {Zoller}(2000)}]{gardiner_quantum_2000}%
  \BibitemOpen
  \bibfield  {author} {\bibinfo {author} {\bibfnamefont {C.~W.}\ \bibnamefont
  {Gardiner}}\ and\ \bibinfo {author} {\bibfnamefont {P.}~\bibnamefont
  {Zoller}},\ }\href@noop {} {\emph {\bibinfo {title} {Quantum {{Noise}}: {{A
  Handbook}} of {{Markovian}} and {{Non}}-{{Markovian Quantum Stochastic
  Methods}} with {{Applications}} to {{Quantum Optics}}}}},\ \bibinfo {edition}
  {2nd}\ ed.\ (\bibinfo  {publisher} {{Springer}},\ \bibinfo {address}
  {{Berlin}},\ \bibinfo {year} {2000})\BibitemShut {NoStop}%
\bibitem [{\citenamefont {Breuer}\ and\ \citenamefont
  {Petruccione}(2002)}]{breuer_the_theory_2002}%
  \BibitemOpen
  \bibfield  {author} {\bibinfo {author} {\bibfnamefont {H.-P.}\ \bibnamefont
  {Breuer}}\ and\ \bibinfo {author} {\bibfnamefont {F.}~\bibnamefont
  {Petruccione}},\ }\href@noop {} {\emph {\bibinfo {title} {The Theory of Open
  Quantum Systems}}},\ \bibinfo {edition} {1st}\ ed.\ (\bibinfo  {publisher}
  {{Oxford University Press}},\ \bibinfo {address} {{New York}},\ \bibinfo
  {year} {2002})\BibitemShut {NoStop}%
\bibitem [{\citenamefont {Giovannetti}\ and\ \citenamefont
  {Vitali}(2001)}]{giovannetti_phase_2001}%
  \BibitemOpen
  \bibfield  {author} {\bibinfo {author} {\bibfnamefont {V.}~\bibnamefont
  {Giovannetti}}\ and\ \bibinfo {author} {\bibfnamefont {D.}~\bibnamefont
  {Vitali}},\ }\href {\doibase 10.1103/PhysRevA.63.023812} {\bibfield
  {journal} {\bibinfo  {journal} {Phys. Rev. A}\ }\textbf {\bibinfo {volume}
  {63}},\ \bibinfo {pages} {023812} (\bibinfo {year} {2001})}\BibitemShut
  {NoStop}%
\bibitem [{\citenamefont {Clerk}\ \emph {et~al.}(2010)\citenamefont {Clerk},
  \citenamefont {Devoret}, \citenamefont {Girvin}, \citenamefont {Marquardt},\
  and\ \citenamefont {Schoelkopf}}]{clerk_introduction_2010}%
  \BibitemOpen
  \bibfield  {author} {\bibinfo {author} {\bibfnamefont {A.~A.}\ \bibnamefont
  {Clerk}}, \bibinfo {author} {\bibfnamefont {M.~H.}\ \bibnamefont {Devoret}},
  \bibinfo {author} {\bibfnamefont {S.~M.}\ \bibnamefont {Girvin}}, \bibinfo
  {author} {\bibfnamefont {F.}~\bibnamefont {Marquardt}}, \ and\ \bibinfo
  {author} {\bibfnamefont {R.~J.}\ \bibnamefont {Schoelkopf}},\ }\href
  {\doibase 10.1103/RevModPhys.82.1155} {\bibfield  {journal} {\bibinfo
  {journal} {Rev. Mod. Phys.}\ }\textbf {\bibinfo {volume} {82}},\ \bibinfo
  {pages} {1155} (\bibinfo {year} {2010})}\BibitemShut {NoStop}%
\bibitem [{\citenamefont {Boventer}\ \emph {et~al.}(2018)\citenamefont
  {Boventer}, \citenamefont {Pfirrmann}, \citenamefont {Krause}, \citenamefont
  {Sch\"on}, \citenamefont {Kl\"aui},\ and\ \citenamefont
  {Weides}}]{Boventer_Complex_2018}%
  \BibitemOpen
  \bibfield  {author} {\bibinfo {author} {\bibfnamefont {I.}~\bibnamefont
  {Boventer}}, \bibinfo {author} {\bibfnamefont {M.}~\bibnamefont {Pfirrmann}},
  \bibinfo {author} {\bibfnamefont {J.}~\bibnamefont {Krause}}, \bibinfo
  {author} {\bibfnamefont {Y.}~\bibnamefont {Sch\"on}}, \bibinfo {author}
  {\bibfnamefont {M.}~\bibnamefont {Kl\"aui}}, \ and\ \bibinfo {author}
  {\bibfnamefont {M.}~\bibnamefont {Weides}},\ }\href {\doibase
  10.1103/PhysRevB.97.184420} {\bibfield  {journal} {\bibinfo  {journal} {Phys.
  Rev. B}\ }\textbf {\bibinfo {volume} {97}},\ \bibinfo {pages} {184420}
  (\bibinfo {year} {2018})}\BibitemShut {NoStop}%
\bibitem [{Note1()}]{Note1}%
  \BibitemOpen
  \bibinfo {note} {Where we adopt the self-energy notation as has been done in
  optomechanics due to the close analogy with how Dyson's equation modifies the
  bare Green's function due to interactions.}\BibitemShut {Stop}%
\end{thebibliography}%

\end{document}